\documentclass[journal,10pt]{IEEEtran}
\usepackage{amsmath,soul}
\usepackage{hyperref}
\usepackage{bm}
\usepackage{algorithm}
\usepackage{epstopdf}
\usepackage{graphicx,epsfig,balance} 
\graphicspath{{./figures/}}
\usepackage[font=small,skip=5pt]{caption}  
\usepackage{cite}
\usepackage{color}
\usepackage{multirow}
\usepackage{amsfonts,amssymb}
\usepackage{diagbox}
\usepackage[noend]{algpseudocode}
\usepackage{booktabs}
\usepackage{threeparttable}
\usepackage{diagbox}
\usepackage[noend]{algpseudocode}
\usepackage{algorithmicx,algorithm}
\usepackage{footnote}
\usepackage{float}

\usepackage{diagbox}
\usepackage{verbatim}
\usepackage{tabularx}
\usepackage{threeparttable}
\usepackage{makecell}
\usepackage{subcaption}
\usepackage{graphicx}
\begin{document}
\title{An Attention-Enhanced $\Phi$-OTDR Event Recognition Framework for Edge-Based Distributed Acoustic Sensing}
\author{
Xiyang Lan, \IEEEmembership{Graduate Student Member, IEEE}, Xin Li, and Yinglei Teng, \IEEEmembership{Senior Member, IEEE}
 \IEEEcompsocitemizethanks{
\IEEEcompsocthanksitem Xiyang Lan and Yinglei Teng are with the Beijing Key Laboratory of Work Safety Intelligent Monitoring, Beijing University of Posts and Telecommunications (BUPT), Xitucheng Road No.10, Beijing, China, 100876.  (e-mail: xiyanglan@bupt.edu.cn;lilytengtt@gmail.com).
\IEEEcompsocthanksitem Xin Li is an incoming PHD student with Data Science and Analytics Thrust, Information Hub, Hong Kong University of Science and Technology (Guangzhou), Guangzhou, China (e-mail: lixin2002cn@gmail.com).
}}
	
\maketitle
\thispagestyle{empty}   
\pagestyle{empty} 

\begin{abstract}
Phase-sensitive optical time-domain reflectometry ($\Phi$-OTDR) has emerged as a promising sensing technology in Internet of Things (IoT) infrastructures, enabling large-scale distributed acoustic sensing (DAS) for real-time monitoring at the edge in smart cities, industrial pipelines, and critical infrastructures. However, accurately recognizing events from massive $\Phi$-OTDR data streams remains challenging, as existing deep learning methods either disrupt the inherent spatiotemporal structure of signals or incur prohibitive computational costs, limiting their applicability in resource-constrained edge computing scenarios. To overcome these challenges, we propose a novel STFT-based Attention-Enhanced Convolutional Neural Network (STFT-AECNN), which represents multi-channel time-series data as stacked spectrograms to fully exploit their spatiotemporal characteristics while enabling efficient 2D CNN processing. A Spatial Efficient Attention Module (SEAM) is further introduced to adaptively emphasize the most informative channels, and a joint Cross-Entropy and Triplet loss is adopted to enhance the discriminability of the learned feature space. Extensive experiments on the public BJTU $\Phi$-OTDR dataset demonstrate that STFT-AECNN achieves a peak accuracy of 99.94\% while maintaining high computational efficiency. These results highlight its potential for real-time, scalable, and robust event recognition in edge-based DAS systems, paving the way for reliable and intelligent IoT sensing applications.
\end{abstract}
	
	
\section{Introduction}
\label{sec1}
Phase-sensitive optical time-domain reflectometry ($\Phi$-OTDR), a core technology underpinning most distributed acoustic sensing (DAS) systems, operates by analyzing the phase variations of Rayleigh backscattered light within optical fibers, enabling highly sensitive detection and precise localization of vibrations \cite{wang2020recent}. $\Phi$-OTDR's inherent advantages, including extensive monitoring reach, high sensitivity, and the ability to repurpose existing telecommunication fiber cables as sensing media \cite{ip2022distributed}, make it a cost-effective solution for large-scale real-time monitoring. With the rapid development of Internet of Things (IoT) infrastructures \cite{ahmed2024toward,alnumay2024past,shafique2020internet,chataut2023unleashing,nizetic2020internet,khanna2020internet}, $\Phi$-OTDR-based DAS systems are increasingly recognized as an enabler for IoT-enabled applications requiring efficient processing at the edge, such as smart city surveillance, industrial pipeline monitoring, and critical infrastructure protection.

Despite the maturity of $\Phi$-OTDR hardware, robust and reliable event recognition remains a major bottleneck to broader deployment in IoT scenarios. This is particularly true for $\Phi$-OTDR hardware, where event recognition in complex environments is a significant challenge. In practice, it is difficult to distinguish genuine threats from benign environmental disturbances or normal human activities, often resulting in high false alarm rates \cite{kandamali2022machine}. These challenges are exacerbated in IoT environments. Here, massive sensing data streams must be processed at the edge under tight constraints of limited bandwidth, computing resources, and latency budgets. This reality makes the development of lightweight and efficient models a critical requirement, a point emphasized by numerous recent studies\cite{guo2024ultralight,liu2025lightweight,cai2024toward,wei2023lightweight,zhao2020lightweight,farooq2022lightweight,agarwal2020lightweight,zhu2021lightweight}. Therefore, developing efficient and accurate event classification algorithms is crucial to unlocking the full potential of $\Phi$-OTDR in large-scale IoT sensing systems.

Conventional approaches rely on handcrafted feature extraction in time, frequency, or time-frequency domains, followed by classical classifiers such as Support Vector Machines (SVM) \cite{yang2022hybrid}, K-Nearest Neighbors (KNN) \cite{wang2022improved}, or decision trees. However, this dependence on domain-specific expertise limits robustness and scalability, as incomplete or redundant features can severely compromise recognition performance. More recently, deep learning \cite{kandamali2022machine,venketeswaran2022recent} has shown great potential by automatically extracting discriminative features from raw DAS signals, achieving higher accuracy and robustness while reducing reliance on manual engineering. Convolutional Neural Networks (CNNs) are particularly popular, with 1D-CNNs focusing on temporal features \cite{wang2021recognition,wu2019one,wu2019vibration}, and 2D-CNNs capturing richer joint time–frequency–spatial features by transforming signals into image-like representations \cite{wang2019practical,ma2022mi,lyu2020distributed,sun2021optical,jiang2018event,shi2022event,shi2019event}. Nevertheless, most CNN-based approaches treat multi-channel inputs in a fixed manner, neglecting spatial heterogeneity across sensing channels, which may degrade classification accuracy. In parallel, more complex models based on LSTM and Transformers \cite{jiang2023adaptive,li2024deep,duan2025inter,jiang2024st,jiang2024high,wu2020novel,chen2020disturbance} have been explored to capture long-range dependencies, but their high computational complexity and large parameter counts limit real-time deployment in IoT edge environments.

In this paper, we propose a lightweight neural network  for $\Phi$-OTDR event recognition, termed STFT-AECNN, which synergistically combines a structured spectrogram representation with an Attention-Enhanced CNN optimized for IoT-constrained scenarios. Specifically, we first transform raw multi-channel $\Phi$-OTDR time-series signals into high-resolution stacked spectrograms using the Short-Time Fourier Transform (STFT). These structured inputs preserve spatial, temporal, and frequency information simultaneously, making them well-suited for efficient 2D CNN processing. We then design a Spatial Efficient Attention Module (SEAM) to adaptively emphasize the most informative channels, and employ a joint Cross-Entropy and Triplet loss to enhance the discriminability of the learned feature space. This holistic design enables the framework to achieve high accuracy and robustness while maintaining computational efficiency, making it suitable for edge-based distributed sensing systems. In summary, the main contributions of this paper are highlighted below.
\begin{itemize}
  \item We propose a multi-channel spectrogram representation for $\Phi$-OTDR signals based on STFT, which preserves temporal dynamics, linear frequency structures, and spatial parallelism, providing structured inputs for efficient feature learning by 2D-CNNs.
  \item We design STFT-AECNN, an attention-enhanced CNN architecture that integrates a Spatial Efficient Attention Module (SEAM) to adaptively focus on heterogeneous sensing channels, and employ a joint Cross-Entropy and Triplet loss to improve the discriminability of the learned feature space.
  \item We validate the proposed framework on the public BJTU $\Phi$-OTDR dataset \cite{cao2023open}, demonstrating state-of-the-art performance with a peak accuracy of 99.94\%. More importantly, the proposed method achieves this performance with high computational efficiency, highlighting its suitability for real-time and edge-based DAS deployments.
\end{itemize}

The remainder of this paper is structured as follows. Section~\ref{related_works} reviews existing studies on $\Phi$-OTDR event classification. Section~\ref{methodology} describes the proposed data representation strategy and the STFT-AECNN architecture in detail. In Section~\ref{result}, we present extensive experiments to evaluate the effectiveness of the framework, including ablation studies on its key components, the SEAM attention module and the joint loss function. Furthermore, a comprehensive comparison with several state-of-the-art approaches is provided to highlight the superiority of the proposed method.

\section{Related Works}
\label{related_works}

\subsection{Sequential Models for \texorpdfstring{$\Phi$}{Phi}\text{-OTDR} Time-Series}
A major class of methods for $\Phi$-OTDR event recognition focuses on processing the parallel one-dimensional time-series signals obtained from each spatial sampling point along the fiber. These approaches aim to learn discriminative representations directly from the raw sequences using deep learning models. For example, some studies employ one-dimensional convolutional neural networks (1D-CNNs) to capture local temporal patterns within individual channels, forming preliminary feature embeddings. These embeddings, or sometimes the multi-channel raw signals themselves, are then passed into recurrent neural networks such as Long Short-Term Memory (LSTM) models and their bidirectional variants \cite{jiang2023adaptive,li2024deep,wu2020novel,chen2020disturbance} to capture long-term temporal dependencies. More recently, Transformer networks with self-attention mechanisms \cite{duan2025inter,jiang2024st,jiang2024high} have been adopted to model intricate long-range correlations across both time and space.

Despite their strong representational power, these sequential models face several challenges in practical $\Phi$-OTDR deployments. First, their high model complexity and large parameter counts require vast labeled datasets, which are expensive and difficult to obtain in real-world sensing environments. Training such models on limited data often results in overfitting and poor generalization to unseen events. Second, a mismatch often exists between the sequential modeling paradigm and the intrinsic structure of $\Phi$-OTDR data. Many discriminative features are inherently two-dimensional—such as spatiotemporal correlations across adjacent channels or localized time-frequency signatures. Sequential models may capture temporal dependencies but fail to exploit these structured 2D patterns effectively. Third, computational efficiency is a critical limitation. Deep LSTMs and Transformer-based models impose significant inference latency, which is unsuitable for latency-sensitive IoT applications such as perimeter security or industrial pipeline monitoring. Their high computational costs also restrict scalability by limiting the number of channels and update frequency a system can support. Therefore, a central challenge is to balance recognition accuracy, model complexity, and efficiency to enable reliable, real-time event recognition in edge-based DAS systems.

\subsection{Image-based Models for \texorpdfstring{$\Phi$-OTDR}{Phi-OTDR} Signals}
An alternative line of research converts one-dimensional vibration signals into two-dimensional images for classification by 2D-CNNs. The effectiveness of this approach critically depends on the imaging strategy. For example, methods that directly map spatiotemporal data into a single-channel grayscale image \cite{cao2023open} may conflate information across dimensions and overlook correlations between spatial channels. Similarly, time-domain transformations such as the Markov Transition Field \cite{mei2025markov} require concatenating multi-channel signals into long sequences followed by downsampling, which disrupts spatial coherence and discards important frequency information.

Time-frequency representations are generally more informative because they reveal how energy evolves across both time and frequency. A common strategy is to generate Mel spectrograms from $\Phi$-OTDR signals and use them as inputs to audio-pretrained CNNs such as VGGish \cite{gan2024fused}. While this leverages transfer learning from the audio domain, it imposes a bias that may be inappropriate for physical vibration analysis. The Mel scale emphasizes low frequencies in line with human auditory perception, but key vibration features in $\Phi$-OTDR may reside in higher frequency ranges. As a result, critical discriminative details risk being compressed or lost, limiting classification performance. This mismatch underscores the need for representations specifically tailored to the physics of DAS signals and their deployment in IoT sensing applications.

\subsection{Research Motivation}
Event recognition in $\Phi$-OTDR systems using deep learning still faces several fundamental challenges in both input representation and model design. Existing methods typically suffer from one or more of the following three major limitations:
\begin{itemize}
    \item \textbf{Neglect of Spatial Information:} Many approaches disrupt the inherent parallel spatial structure of $\Phi$-OTDR signals. For example, the aforementioned 1D sequential models often treat spatial channels independently, while certain 2D imaging strategies obscure or distort critical cross-channel correlations.

    \item \textbf{Suboptimal Feature Representation:} Transforming raw signals into images can result in information loss even within a single channel. Specifically, in a common strategy borrowed from the audio domain, transforms such as the audio-oriented Mel scale, suppress high-frequency details that are essential for distinguishing $\Phi$-OTDR events, thereby compromising the fidelity and discriminative power of the input representation.

    \item \textbf{Inefficient or Ineffective Feature Learning:}A difficult trade-off persists between model expressiveness and computational efficiency. On one hand, standard CNNs, while efficient, often treat all channels uniformly, lacking the ability to dynamically emphasize the most salient spatial channels. On the other hand, the more powerful sequential models, such as LSTMs and Transformers, solve for expressiveness but incur prohibitive computational complexity and inference latency, making them impractical for real-time deployment.
\end{itemize}

To overcome these challenges, we address the first two limitations by introducing a multi-channel spectrogram representation. This representation preserves the parallel spatial structure by treating each fiber location as a distinct image channel and employs a linear, high-fidelity frequency mapping without domain-specific bias. The resulting data structure unifies spatial information (across channels) with temporal–frequency dynamics (within each spectrogram), providing both a richer and more faithful foundation for feature learning.

Building upon this optimized input, we address the third limitation  by developing STFT-AECNN, an attention-enhanced CNN architecture. The model incorporates a Spatial Efficient Attention Module (SEAM) that adaptively evaluates the relevance of each channel, allowing the network to focus on the most discriminative information. In this way, STFT-AECNN achieves a favorable balance between accuracy and efficiency—avoiding the complexity of purely sequential models—while ensuring suitability for real-time, large-scale deployment in edge-based distributed acoustic sensing systems.

\section{Our Proposed Method}
\label{methodology}
This section presents our proposed $\Phi$-OTDR event recognition framework, which consists of two key components: (i) a novel multi-channel STFT-based spectrogram representation for signal preprocessing, and (ii) the STFT-AECNN, an attention-enhanced Convolutional Neural Network designed for efficient feature learning and classification.  

\subsection{Overview}
\begin{figure*}[htbp]
    \centering
    \includegraphics[width=\textwidth]{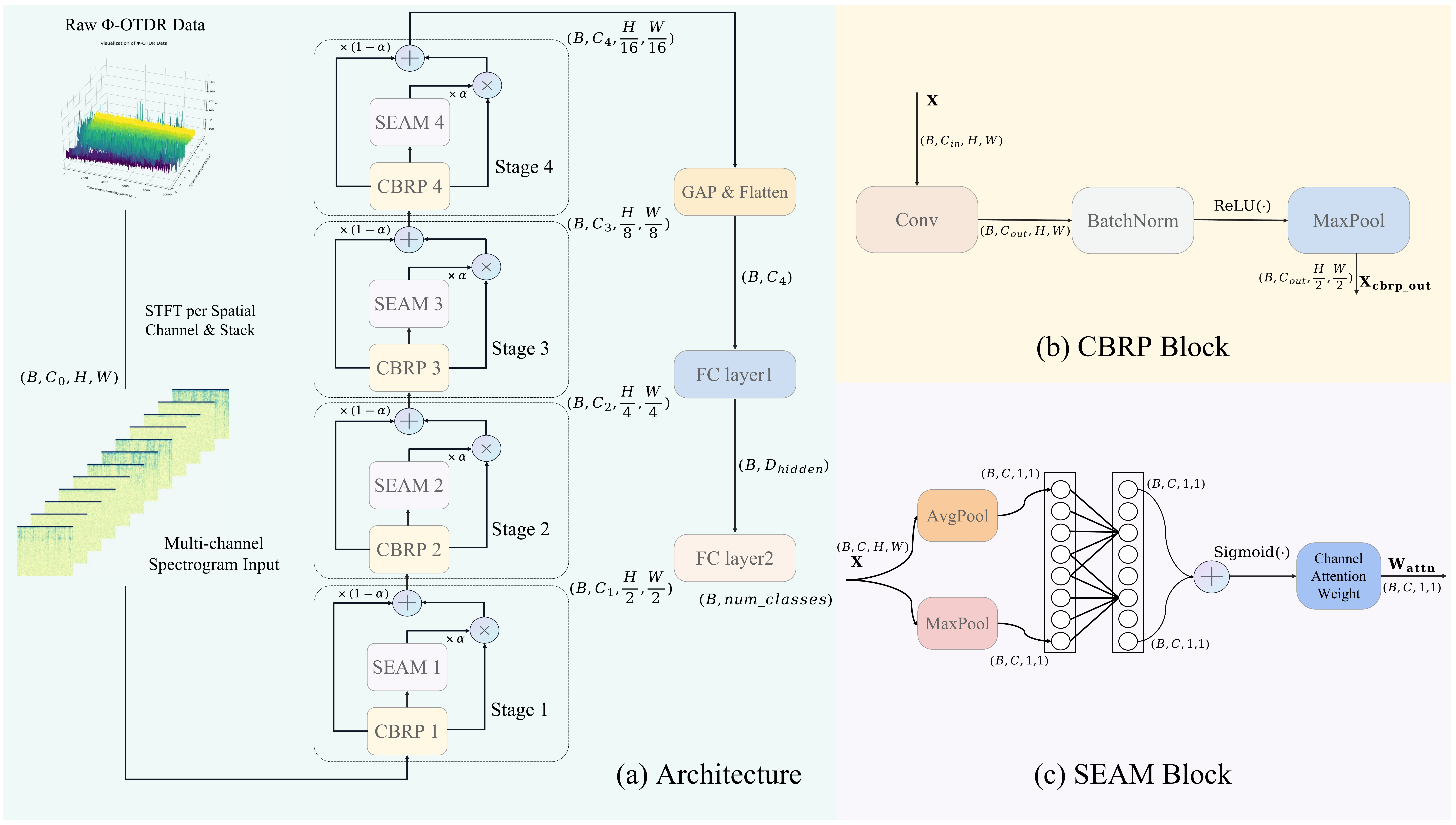}
    \caption{Architecture of the proposed STFT-AECNN. 
    (a) The multi-stage backbone architecture. 
    (b) The Convolutional-BatchNorm-ReLU-MaxPool (CBRP) block. 
    (c) The Spatial Efficient Attention Module (SEAM) block.}
    \label{fig:architecture}
\end{figure*}

The overall architecture of the proposed STFT-AECNN framework is illustrated in Fig.~\ref{fig:architecture}. The framework follows a hierarchical, multi-stage design to progressively extract and refine discriminative features. Our framework first transforms the raw $\Phi$-OTDR signals into multi-channel STFT spectrograms through its preprocessing pipeline. These spectrograms serve as structured, information-rich inputs to the network.  

At the core of the architecture lies an iterative refinement process. Each stage initially extracts local spectral–temporal features from the spectrograms of individual channels. These features are then integrated using a novel attention-based fusion mechanism, which adaptively re-weights the contributions from different spatial channels. In this way, the model dynamically emphasizes the most informative representations while retaining the inherent richness of the data.  After passing through cascaded convolutional and attention stages, the high-level features are aggregated and fed into a classification head to produce the final prediction. This structured approach enables the network to effectively learn complex spatial–spectral representations that are critical for distinguishing diverse event types in $\Phi$-OTDR systems.  

\subsection{Multi-Channel Spectrogram Generation from Raw \texorpdfstring{$\Phi\text{-OTDR}$}{Phi-OTDR} Data}
To fully capture the spatial and frequency characteristics inherent in $\Phi$-OTDR signals, we design a preprocessing pipeline that transforms raw multi-channel time-series data into a stacked spectrogram representation. This transformation is performed independently for each spatial channel, ensuring that the unique time–frequency properties of each location are preserved. The resulting spectrograms are then stacked, creating a unified multi-channel format that serves as the input to the STFT-AECNN.  

Let the raw $\Phi$-OTDR data for a single event be denoted as $\mathbf{X}_{\text{raw}} \in \mathbb{R}^{N_s \times N_t}$, where $N_s$ is the number of spatial sampling points (channels) along the fiber, and $N_t$ is the number of temporal samples per channel. The preprocessing consists of the following steps for each spatial channel $i$, where $i = 1, 2, \cdots, N_s$:  

\begin{itemize}
    \item \textbf{Signal Selection:}  
    For each event, the raw time-series signal from the $i$-th spatial channel is denoted as $\mathbf{x}_i[n] \in \mathbb{R}^{N_t}$.  

    \item \textbf{Short-Time Fourier Transform (STFT):}  
    The STFT is applied to each signal to obtain its time–frequency representation. For a continuous signal $\mathbf{x}_i(t)$, the STFT is defined as:
    \begin{equation}
        S_i(\tau, f) = \int_{-\infty}^{\infty} \mathbf{x}_i(t) w(t-\tau) e^{-j2\pi ft} dt ,
    \end{equation}
    where $w(t)$ is a window function. In practice, we compute the discrete STFT using the DFT over windowed frames of $\mathbf{x}_i[n]$. For the $m$-th frame at frequency bin $k$, this is expressed as:
    \begin{equation}
        S_i[m, k] = \sum_{n=0}^{L-1} \mathbf{x}_i[n+mH] w[n] e^{-j\frac{2\pi k n}{N}},
    \end{equation}

    where $L$ is the window length, $H$ is the hop length, and $N$ is the FFT size. The FFT algorithm is used for efficient computation, and parameters are chosen to achieve the desired time–frequency resolution.  
The process of applying a sliding window to a single-channel signal is visualized in Fig.~\ref{fig:stft_process}, which, for clarity, displays only the initial segment of the time-series to illustrate how the signal is segmented into overlapping frames.
\begin{figure}[htbp]
    \centering
    \includegraphics[width=0.48\textwidth]{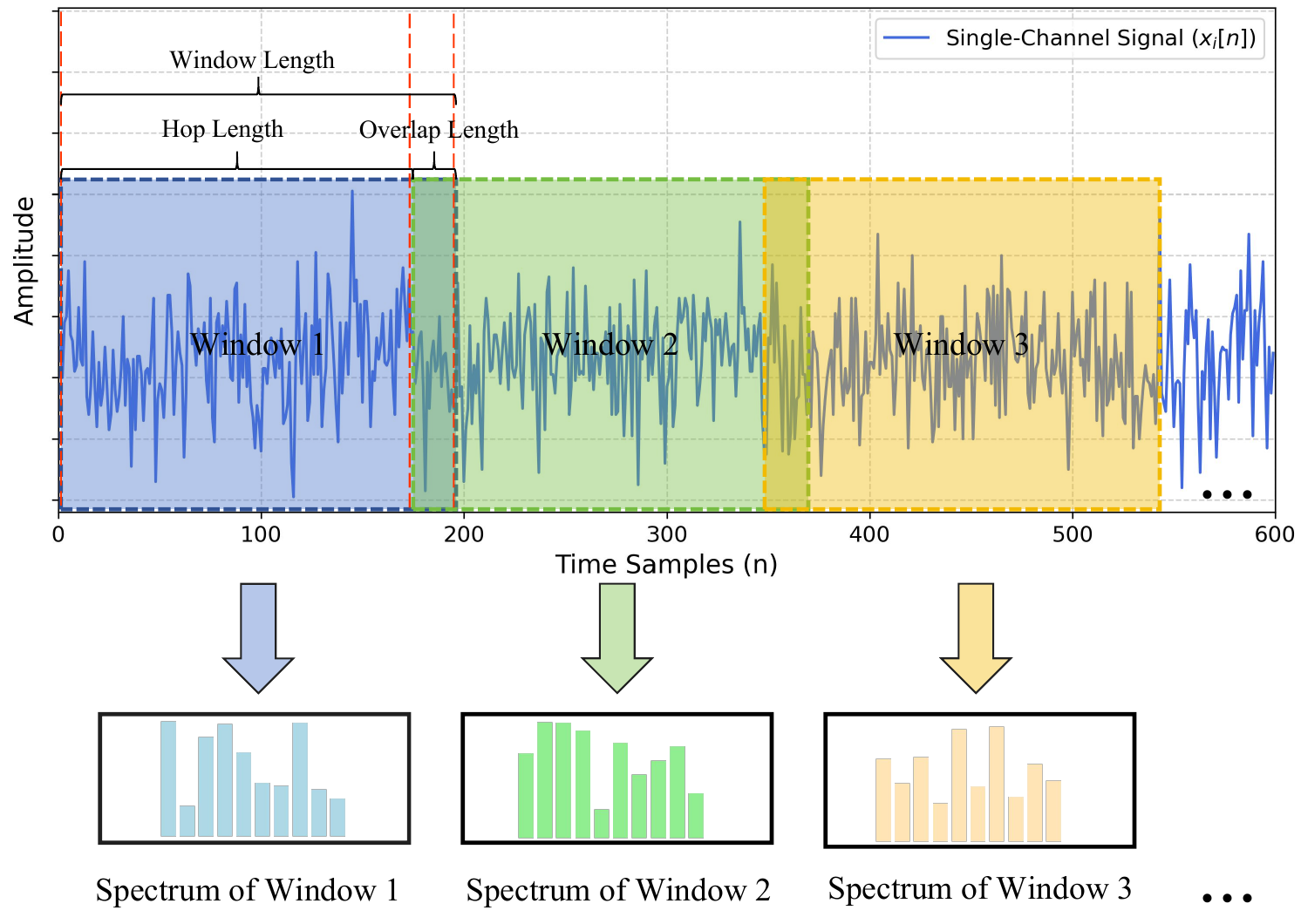} 
    \caption{The STFT process: from time-series to spectrogram. A single-channel signal is partitioned into overlapping windows, each of which is converted into a spectral column via FFT to construct the spectrogram.}
    \label{fig:stft_process}
\end{figure}

    \item \textbf{Spectrogram Post-processing:}  
    The magnitude spectrogram is obtained as
    \begin{equation}
        \mathbf{P}_i[k, m] = |S_i[m, k]| ,
    \end{equation}
    where $k$ and $m$ denote frequency-bin and time-frame indices, respectively. To improve stability, logarithmic compression ($f(x) = \log(1+x)$) is applied, followed by min–max normalization to scale the values to [0, 1]. The result is a single-channel image $\mathbf{P}_i \in \mathbb{R}^{H \times W}$.  

    \item \textbf{Multi-Channel Stacking:}  
    After obtaining spectrograms for all $N_s$ channels, they are stacked along a new dimension to form the final multi-channel input:  
    \begin{equation}
        \mathbf{X}_{\text{in}} \in \mathbb{R}^{N_s \times H \times W}.
    \end{equation}
    This preserves both the spatial channel structure and the time–frequency features, providing a rich and well-structured representation for subsequent learning.
\end{itemize}

To provide a clear and procedural overview, the complete data preprocessing pipeline is summarized in Algorithm~\ref{alg:spectrogram_generation}, which details the transformation from raw time-series data into the final multi-channel spectrogram tensor.

\begin{algorithm}[h!]
\caption{Multi-Channel Spectrogram Generation}
\label{alg:spectrogram_generation}
\begin{algorithmic}[1]
\Require Raw $\Phi$-OTDR data matrix $\mathbf{X}_{\text{raw}} \in \mathbb{R}^{N_s \times N_t}$ \\
         STFT parameter set \texttt{params}: \texttt{window\_type}, \texttt{win\_length}, \texttt{hop\_length}
\Ensure Multi-channel spectrogram tensor $\mathbf{X}_{\text{in}} \in \mathbb{R}^{N_s \times H \times W}$

\Function{GenerateSpectrogram}{$\mathbf{X}_{\text{raw}}$, \texttt{params}}
    \State Initialize empty list \texttt{spectrogram\_list} $\leftarrow []$
    \For{$i \leftarrow 1$ \textbf{to} $N_s$} \Comment{Process each spatial channel}
        \State \textit{// Step 1: Signal Selection}
        \State $\mathbf{x}_i \leftarrow \text{GetRow}(\mathbf{X}_{\text{raw}}, i)$
        \State \textit{// Step 2: Short-Time Fourier Transform (STFT)}
        \State $S_i \leftarrow \text{STFT}(\mathbf{x}_i, \texttt{params})$
        \State \textit{// Step 3: Magnitude and Post-processing}
        \State $\mathbf{P}_i \leftarrow |S_i|$
        \State $\mathbf{P}_i \leftarrow \log(1 + \mathbf{P}_i)$
        \State Normalize $\mathbf{P}_i$ to $[0, 1]$
        \State Append $\mathbf{P}_i$ to \texttt{spectrogram\_list}
    \EndFor
    \State \textit{// Step 4: Multi-Channel Stacking}
    \State $\mathbf{X}_{\text{in}} \leftarrow \text{Stack}(\texttt{spectrogram\_list})$
    \State \Return $\mathbf{X}_{\text{in}}$
\EndFunction
\end{algorithmic}
\end{algorithm}

This preprocessing pipeline transforms parallel one-dimensional time-series signals into a unified multi-channel spectrogram tensor. Each channel of the tensor corresponds to the time–frequency representation of a specific spatial location, while the spectrogram plane captures the temporal evolution of frequency components for that location. This structure enables our model to simultaneously learn (i) local time–frequency patterns within each channel, and (ii) cross-channel correlations and discriminative spatial variations through multi-channel convolutions and subsequent attention mechanisms.  

\begin{figure*}[htbp]
    \centering
    \includegraphics[width=\textwidth]{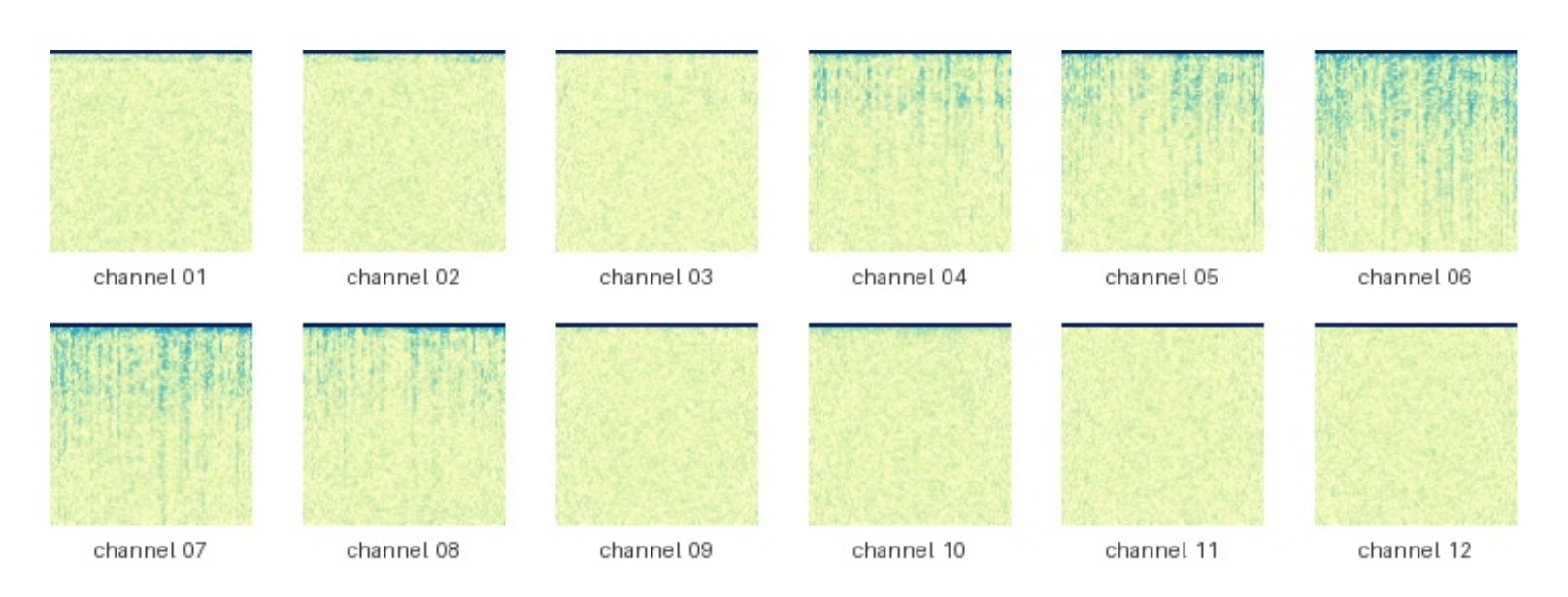}
    \caption{Example of a 12-channel spectrogram for a single \textit{watering} event sample. Discriminative features are localized in channels 4–8, highlighting the importance of adaptive channel selection.}
    \label{fig:watering_sample}
\end{figure*}

As illustrated in Fig.~\ref{fig:watering_sample}, the discriminative patterns of a sample from the \textit{watering} category—characterized by dense, waterfall-like textures caused by broadband droplet noise—are concentrated in channels 4 through 8, while other channels contain mostly background noise. This spatial heterogeneity underscores the need for a mechanism that can dynamically identify and focus on the most informative channels. Our attention-enhanced CNN is explicitly designed for this purpose.

\subsection{Proposed Network Architecture}
\subsubsection{Overall Architecture and Data Flow}
\label{subsec:overall_architecture}
As illustrated in Fig.~\ref{fig:architecture}(a), STFT-AECNN adopts a hierarchical four-stage design to process multi-channel STFT spectrograms, where each input channel corresponds to a distinct spatial sensing location. Each stage employs a dual-path attention–fusion structure.  

First, the input passes through a Convolutional-BatchNorm-ReLU-MaxPool (CBRP) block (detailed in Section~\ref{subsubsec:cbrp}) to extract local features. Then, the output of the CBRP is split into two parallel paths: one path is processed by the Spatial Efficient Attention Module (SEAM, see Section~\ref{subsubsec:seam}) for dynamic channel-wise recalibration, while the other retains the original features through a residual skip connection.  

The two paths are fused using a learnable scalar parameter $\alpha$, which adaptively balances attention-enhanced and residual features:
\begin{equation}
y_{\mathrm{stage}} = \alpha \cdot \mathrm{SEAM}(\mathrm{CBRP}_{\mathrm{out}}) + (1 - \alpha) \cdot \mathrm{CBRP}_{\mathrm{out}} .
\label{eq:fusion}
\end{equation}
This fusion mechanism allows the model to dynamically control the extent of attention modulation at each stage. Repeating this process across the four stages yields progressively abstract representations.  

The final feature maps are aggregated via Global Average Pooling (GAP), flattened, and passed into a two-layer fully connected classification head. The first layer applies ReLU activation and Dropout regularization, while the second produces logits for the $N_c$ event categories. The entire model is optimized under the joint loss function described in Section~\ref{subsubsec:loss_function}.
\subsubsection{CBRP Module} 
\label{subsubsec:cbrp}
The CBRP module, illustrated in Fig.~\ref{fig:architecture}(b), serves as the primary unit for local feature extraction within each stage. It sequentially applies four operations to its input feature map, $\mathbf{X}_{\mathrm{in}}$.

First, a 2D convolutional layer with $C_{\mathrm{cbrp\_out}}$ filters detects local spectral-temporal patterns. Its $k$-th output channel, $\mathbf{X}_{\mathrm{conv}}^{(k)}$, is given by:
\begin{equation}
\mathbf{X}_{\mathrm{conv}}^{(k)} = \sum_{c=1}^{C_{\mathrm{in}}} \left( \mathbf{X}_{\mathrm{in}}^{(c)} * \mathbf{W}_{\mathrm{conv}}^{(k,c)} \right) + \mathbf{b}_{\mathrm{conv}}^{(k)}
\label{eq:convolution}
\end{equation}
The feature map is then processed through a sequence of normalization, activation, and pooling layers. First, it is stabilized by a Batch Normalization (BN) layer. This is followed by a Rectified Linear Unit (ReLU) to introduce element-wise non-linearity. These operations can be expressed as:
\begin{align}
    \mathbf{X}_{\mathrm{bn}} &= \mathrm{BN}(\mathbf{X}_{\mathrm{conv}}; \gamma, \beta) \\
    \mathbf{X}_{\mathrm{relu}} &= \mathrm{ReLU}(\mathbf{X}_{\mathrm{bn}}) = \max(0, \mathbf{X}_{\mathrm{bn}})
\end{align}
where $\gamma$ and $\beta$ are the learnable scaling and shifting parameters of the BN layer.
\par 
Finally, a Max Pooling layer performs spatial downsampling with a $K_p \times K_p$ window and a stride of $S_p$, reducing the feature map's spatial dimensions while preserving the $C_{\mathrm{cbrp\_out}}$ channels. The final output of the module, $\mathbf{X}_{\mathrm{cbrp\_out}}$, is thus given by:
\begin{equation}
    \mathbf{X}_{\mathrm{cbrp\_out}} = \mathrm{MaxPool}(\mathbf{X}_{\mathrm{relu}})
\end{equation}

\subsubsection{The SEAM Module} 
\label{subsubsec:seam}
The SEAM, depicted in Fig.~\ref{fig:architecture}(c), is designed to perform efficient channel attention. It takes the feature map from the preceding CBRP module and adaptively recalibrates the importance of each channel to enhance feature representation.

First, the module aggregates spatial information into two distinct channel descriptors via parallel pooling: Global Average Pooling, yielding $\mathbf{P}_{\mathrm{avg}}$, and Global Max Pooling, yielding $\mathbf{P}_{\mathrm{max}}$. These descriptors are then processed independently by a shared 1D convolutional layer. The kernel size of this layer, $k_{\mathrm{eca}}$, is adaptively determined from the number of input channels, $C_{\mathrm{cbrp\_out}}$, using the following relationship:
\begin{equation}
    k_{\mathrm{eca}} = \left| \frac{\log_2(C_{\mathrm{cbrp\_out}}) + b}{\gamma} \right|_{\mathrm{odd}}
    \label{eq:k_eca} 
\end{equation}
with typical defaults being $\gamma=2$ and $b=1$. The processed features from both pooling paths, $\mathbf{V}_{\mathrm{avg}}$ and $\mathbf{V}_{\mathrm{max}}$, are then element-wise added. The sum is passed through a Sigmoid activation function to generate the final channel attention weights, $\mathbf{W}_{\mathrm{attn}}$:
\begin{equation}
    \mathbf{W}_{\mathrm{attn}} = \sigma(\mathbf{V}_{\mathrm{avg}} + \mathbf{V}_{\mathrm{max}})
    \label{eq:attention_weights}
\end{equation}
Finally, the original input feature map $\mathbf{X}_{\mathrm{cbrp\_out}}$ is recalibrated by element-wise multiplication with the computed attention weights, producing the module's output, $\mathbf{X}_{\mathrm{seam\_out}}$, as defined by:
\begin{equation}
    \mathbf{X}_{\mathrm{seam\_out}} = \mathbf{X}_{\mathrm{cbrp\_out}} \odot \mathbf{W}_{\mathrm{attn}}
    \label{eq:seam_output} 
\end{equation}
This attention-refined feature map, which serves as an input to the subsequent alpha-weighted fusion mechanism, retains its original dimensions but with its channel features adaptively re-weighted. Substituting this back into the stage's fusion equation (Eq.~\ref{eq:fusion}), the complete expression for a stage's output is:
\begin{equation}
    \mathbf{y}_{\mathrm{stage}} = \alpha (\mathbf{X}_{\mathrm{cbrp\_out}} \odot \mathbf{W}_{\mathrm{attn}}) + (1 - \alpha)  \mathbf{X}_{\mathrm{cbrp\_out}}.
    \label{eq:stage_output_full}  
\end{equation}
\subsubsection{Joint Loss Function for Optimization}
\label{subsubsec:loss_function}

To train STFT-AECNN effectively for both accurate event classification and the learning of discriminative feature embeddings, we adopt a joint loss function, $\mathcal{L}_{\text{total}}$. This composite objective integrates a standard classification loss with a metric learning loss, thereby guiding optimization from two complementary perspectives.

The first component is the \textbf{Cross-Entropy (CE) Loss}, denoted as $\mathcal{L}_{\text{CE}}$, which is widely used in multi-class classification tasks. It measures the divergence between the predicted probability distribution and the ground-truth label. Given output logits $\mathbf{p} \in \mathbb{R}^{N_c}$ for $N_c$ classes and the true label index $y_{\text{true\_idx}}$, the CE loss is expressed as:
\begin{equation}
\mathcal{L}_{\text{CE}}(\mathbf{p}, y_{\text{true\_idx}}) = -\log\left(\frac{\exp(p_{y_{\text{true\_idx}}})}{\sum_{j=1}^{N_c} \exp(p_j)}\right).
\label{eq:cross_entropy_loss}
\end{equation}
This term provides direct supervision for the classification head.

The second component is the \textbf{Triplet Loss}, denoted as $\mathcal{L}_{\text{Triplet}}$, which acts on the feature embeddings $\mathbf{emb}$ extracted by the backbone before the final classification layers. Its objective is to enforce intra-class compactness and inter-class separability in the embedding space. For an anchor embedding $\mathbf{emb}_a$, a positive embedding $\mathbf{emb}_p$ (same class), and a negative embedding $\mathbf{emb}_n$ (different class), the triplet loss for one triplet is defined as:
\begin{equation}
\begin{split}
\mathcal{L}_{\text{triplet\_single}} = \max(0, & D(\mathbf{emb}_a, \mathbf{emb}_p)^2 \\
                                               & - D(\mathbf{emb}_a, \mathbf{emb}_n)^2 + m),
\end{split}
\label{eq:your_label_for_triplet_loss}
\end{equation}
where $D(\cdot, \cdot)$ denotes the Euclidean distance and $m$ is a margin parameter. The batch-level $\mathcal{L}_{\text{Triplet}}$ is computed as the average over all valid triplets.

Finally, the total loss for optimization is formulated as:
\begin{equation}
\mathcal{L}_{\text{total}} = \mathcal{L}_{\text{CE}} + \mathcal{L}_{\text{Triplet}}.
\label{eq:total_loss}
\end{equation}

By minimizing $\mathcal{L}_{\text{total}}$, STFT-AECNN is trained not only to achieve high classification accuracy via $\mathcal{L}_{\text{CE}}$, but also to learn a structured and discriminative feature space through $\mathcal{L}_{\text{Triplet}}$. This combination is particularly beneficial for distinguishing subtle variations in $\Phi$-OTDR events, where different classes may exhibit fine-grained similarities.

\section{Experimental Results and Analysis}
\label{result}
This section presents a comprehensive evaluation of the proposed attention-enhanced CNN framework for $\Phi$-OTDR event recognition. We begin by describing the experimental settings, including implementation details, dataset characteristics, and evaluation metrics. We then conduct ablation studies to assess the effectiveness of key architectural components, namely the attention mechanism and the joint loss function. Finally, we provide a comparative analysis against several representative baselines to demonstrate the superiority of our approach.

\subsection{Experimental Settings}
\label{sec:experimental_settings}
\noindent \textbf{Implementation Details:}  
All experiments were implemented in Python 3.8 (Ubuntu 20.04) using PyTorch 1.10.0 with CUDA 11.3 acceleration. Training and evaluation were performed on a high-performance computing server equipped with an NVIDIA GeForce RTX 4090 GPU (24\,GB VRAM), a 16-core Intel Xeon Gold 6430 CPU, and 120\,GB of RAM.

\begin{figure}[t!]
    \centering
    \includegraphics[width=0.95\columnwidth]{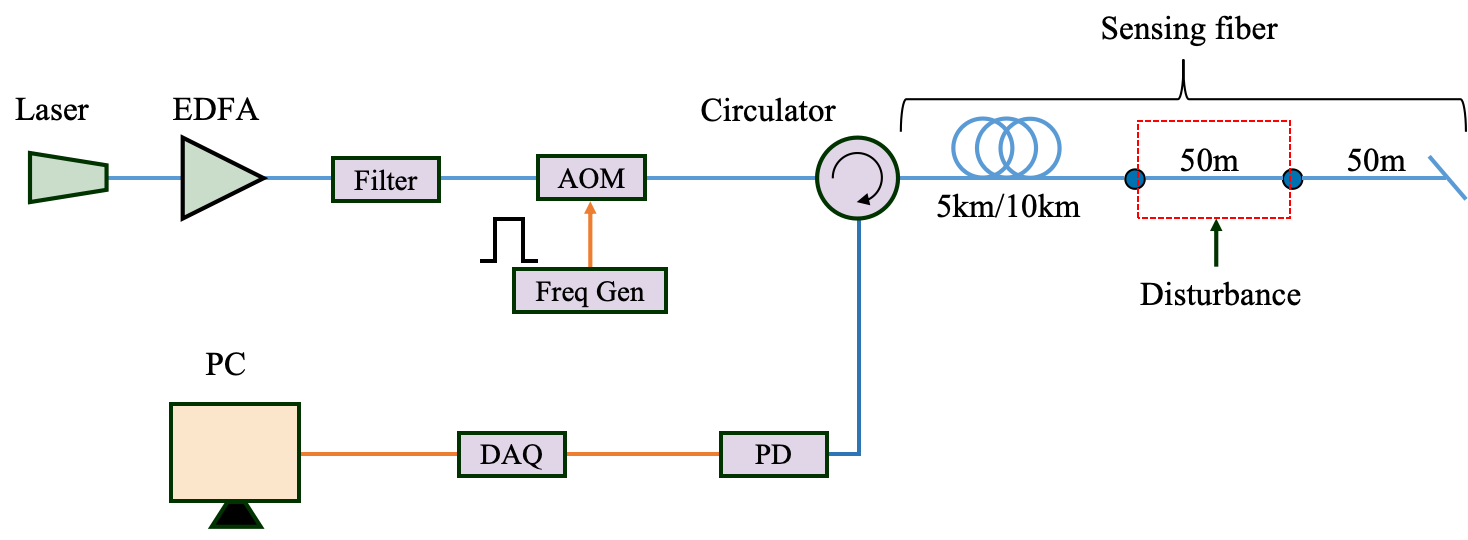}
    \caption{Schematic of the experimental setup for $\Phi$-OTDR dataset acquisition.}
    \label{fig:experimental_setup}
\end{figure}

\noindent \textbf{Dataset:}  
We used the publicly available $\Phi$-OTDR event dataset published by Cao \textit{et al.} \cite{cao2023open}, collected with an intensity-based $\Phi$-OTDR system. The experimental setup is illustrated in Fig.~\ref{fig:experimental_setup}. The sensing fiber lengths were 5 km and 10 km, each terminated by a 100 m armored section where events were induced. All event data were acquired within the first 50 m of the armored fiber. The pulse repetition frequency was set to 12.5 kHz and 8 kHz for the 5 km and 10 km configurations, respectively, while the data acquisition card (DAQ) sampling rate was fixed at 10 MSa/s for both settings.

\begin{table}[h!]
\centering
\caption{Composition of the BJTU $\Phi$-OTDR Event Dataset}
\label{tab:dataset_distribution}
\begin{tabular}{lcr}
\toprule
\textbf{Event Category} & \textbf{Samples} & \textbf{Label} \\
\midrule
Background noise        & 2946  & 0 \\
Digging                 & 2512  & 1 \\
Knocking                & 2530  & 2 \\
Watering                & 2253  & 3 \\
Shaking                 & 2728  & 4 \\
Walking                 & 2450  & 5 \\
\midrule
\textbf{Total}          & \textbf{15419} & -- \\
\bottomrule
\end{tabular}
\end{table}

As shown in Table~\ref{tab:dataset_distribution}, the dataset contains 15,419 samples across six event categories: background noise, digging, knocking, watering, shaking, and walking. Each sample is a spatiotemporal matrix of dimension $10,000 \times 12$, corresponding to temporal samples across 12 adjacent spatial channels. Following the protocol of \cite{cao2023open}, the dataset was randomly split into training and testing sets in an 8:2 ratio, ensuring no overlap.

\begin{figure}[htbp]
    \centering
    \includegraphics[
        width=0.7\textwidth,
        height=0.5\textheight,
        keepaspectratio
    ]{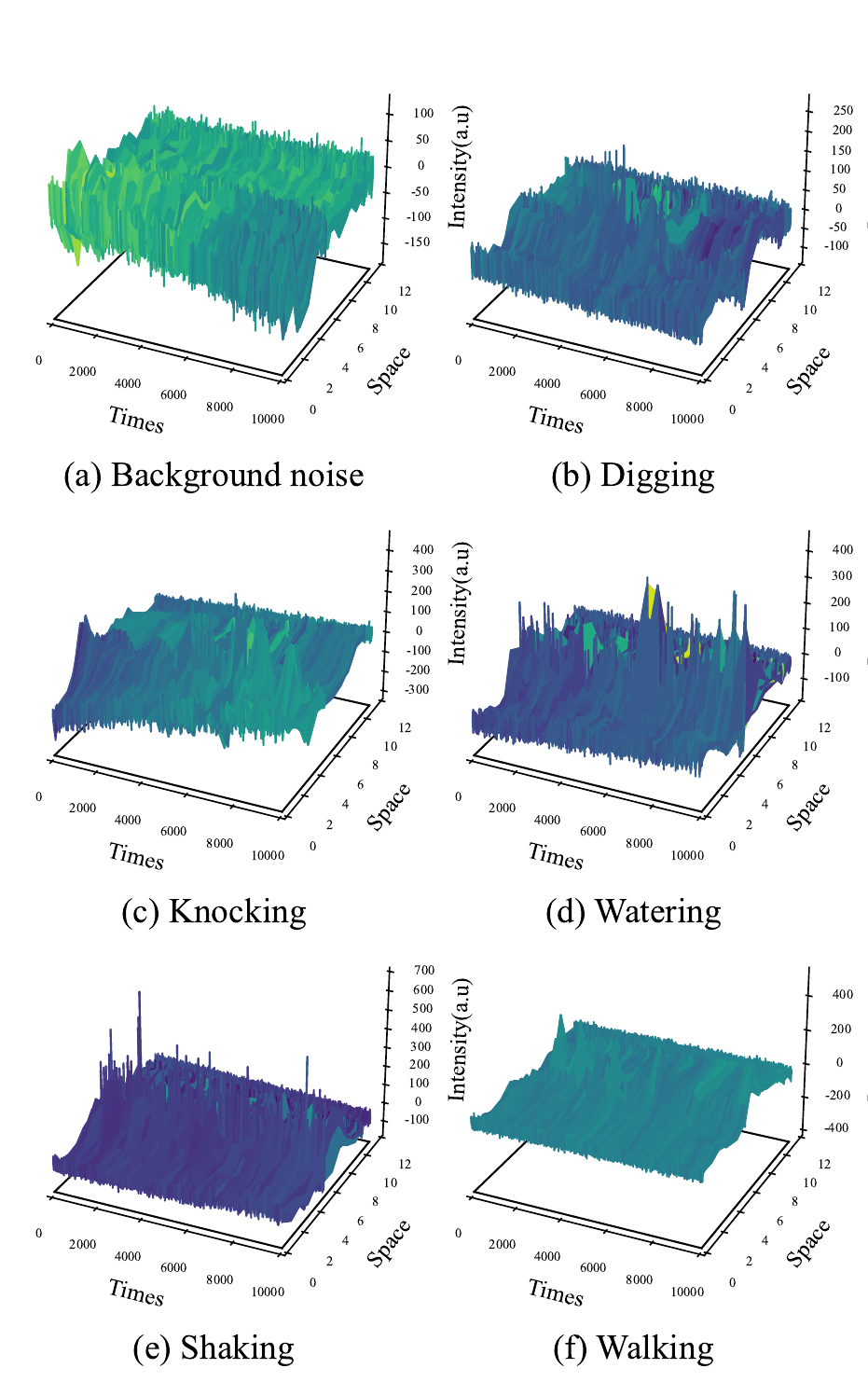} 
    \caption{Spatiotemporal visualizations of six representative event types: (a) Background noise, (b) Digging, (c) Knocking, (d) Watering, (e) Shaking, and (f) Walking.}
    \label{fig:das_events}
\end{figure}

The six event types exhibit distinct spatiotemporal and spectral signatures. Background noise (a) appears as low-amplitude random fluctuations, while impulsive events such as digging (b) and knocking (c) produce localized, high-energy peaks spread across a wide frequency band. Shaking (e) is characterized by an initial impact followed by sustained oscillations, yielding both broadband and high-frequency components. In contrast, long-duration activities such as watering (d) and walking (f) manifest as smoother temporal patterns with energy concentrated in low-frequency ranges. In particular, walking (f) produces a broad, continuous disturbance consistent with cumulative ground vibrations from footsteps. These complementary patterns in both temporal and spectral domains provide a robust basis for discriminative classification.

\noindent \textbf{Parameter Selection:}  
For preprocessing, raw time-series signals were transformed into spectrograms using STFT, as described in Section~\ref{methodology}. Each of the 12 spatial channels was independently transformed, yielding a tensor of size $12 \times 100 \times 100$ per event. The STFT parameters are summarized in Table~\ref{tab:stft_params}.

\begin{table}[htbp]
\centering
\caption{Parameters for STFT-based Spectrogram Generation}
\label{tab:stft_params}
\begin{tabular}{lc}
\hline
\textbf{Parameter} & \textbf{Value} \\
\hline
Input Signal Length (per channel) & 10000 \\
Target Spectrogram Dimension & $100 \times 100$ \\
Window Type & Hann \\
Window Length ($L$) & 198 \\
FFT Size ($N$) & 198 \\
Hop Length ($H$) & 100 \\
Overlap Length & 98 \\
\hline
\end{tabular}
\end{table}

\noindent \textbf{Training Protocol:}  
The proposed network was implemented following the architecture described in Section~\ref{methodology}. Training was conducted using the Adam optimizer with an initial learning rate of $1 \times 10^{-4}$ and weight decay of $1 \times 10^{-5}$. The joint loss function combined Cross-Entropy Loss with a Triplet Loss (margin $m=0.0$). Models were trained for 50 epochs with a batch size of 50. To ensure reproducibility, a fixed random seed was applied throughout all experiments.

\subsection{Evaluation Metrics}

To evaluate our model's classification performance, we use standard metrics derived from the Confusion Matrix, which provides counts for True Positives (TP), True Negatives (TN), False Positives (FP), and False Negatives (FN).

Based on these counts, we calculate four key metrics. The overall performance is measured by \textbf{Accuracy}, which reflects the proportion of all correct predictions:
\begin{equation}
    \text{Accuracy} = \frac{\text{TP} + \text{TN}}{\text{TP} + \text{TN} + \text{FP} + \text{FN}}
    \label{eq:accuracy}
\end{equation}
To analyze performance on specific event types, we utilize \textbf{Precision} and \textbf{Recall}. Precision measures the exactness of the classification, representing the ratio of correctly predicted positive samples to all predicted positive samples:
\begin{equation}
    \text{Precision} = \frac{\text{TP}}{\text{TP} + \text{FP}}
    \label{eq:precision}
\end{equation}
Recall measures the completeness of the classification, representing the ratio of correctly predicted positive samples to all actual positive samples:
\begin{equation}
    \text{Recall} = \frac{\text{TP}}{\text{TP} + \text{FN}}
    \label{eq:recall}
\end{equation}
The \textbf{F1-Score}, the harmonic mean of Precision and Recall, serves as a balanced metric for performance:
\begin{equation}
    \text{F1-Score} = \frac{2 \cdot \text{TP}}{2 \cdot \text{TP} + \text{FP} + \text{FN}}
    \label{eq:f1_score}
\end{equation}
In our multi-class evaluation, we report the overall accuracy as a top-level metric and present per-class Precision, Recall, and F1-score for a more granular analysis.

	\subsection{Ablation Experiment}
	\label{sec:ablation}
To validate the individual contributions of the key components within our proposed framework, we conduct a comprehensive ablation study. This study compares the performance of  STFT-AECNN against two ablated variants:
\begin{itemize}
    \item \textbf{STFT-AECNN:} The complete proposed architecture, incorporating both the SEAM attention mechanism and the joint loss function.
    \item \textbf{w/o SEAM Attention:} The model with the SEAM attention modules removed from each stage.
    \item \textbf{w/o Triplet Loss:} The model trained solely with Cross-Entropy Loss, excluding the triplet loss component from the joint loss function.
\end{itemize}

To ensure a fair comparison, all model configurations were trained following the same experimental settings detailed in Section~\ref{sec:experimental_settings}. We evaluate each variant based on its overall classification accuracy and a detailed analysis of its per-class performance and error types, as revealed by the confusion matrix.
The results of our ablation study are presented in Table~\ref{tab:ablation_study}.  STFT-AECNN achieves a peak accuracy of 99.94\%, which serves as the baseline for evaluating the ablated models. The study clearly validates the contributions of each key component. As shown in Table~\ref{tab:ablation_study}, removing the SEAM attention module results in a 0.17\% accuracy degradation, confirming its vital role in enhancing feature discriminability. Similarly, excluding the Triplet Loss leads to a 0.13\% performance drop, which underscores its importance for learning a robust feature embedding space. These findings demonstrate that the synergy between both components is crucial for achieving this state-of-the-art performance.

\begin{table}[htbp]
    \centering
    \caption{Results of the Ablation Study} 
    \label{tab:ablation_study}
    \begin{tabular}{lcc}
        \toprule
        \textbf{Model Configuration} & \textbf{Peak Accuracy (\%)} & \textbf{Accuracy Drop (\%)} \\  
        \midrule
        STFT-AECNN              & 99.94 & - \\  
        w/o SEAM Attention      & 99.77 & 0.17 \\  
        w/o Triplet Loss        & 99.81 & 0.13 \\  
        \bottomrule
    \end{tabular}
\end{table}

\begin{figure}[htbp]
    \centering 

    \begin{minipage}{0.48\columnwidth}
        \centering
        \includegraphics[width=\textwidth]{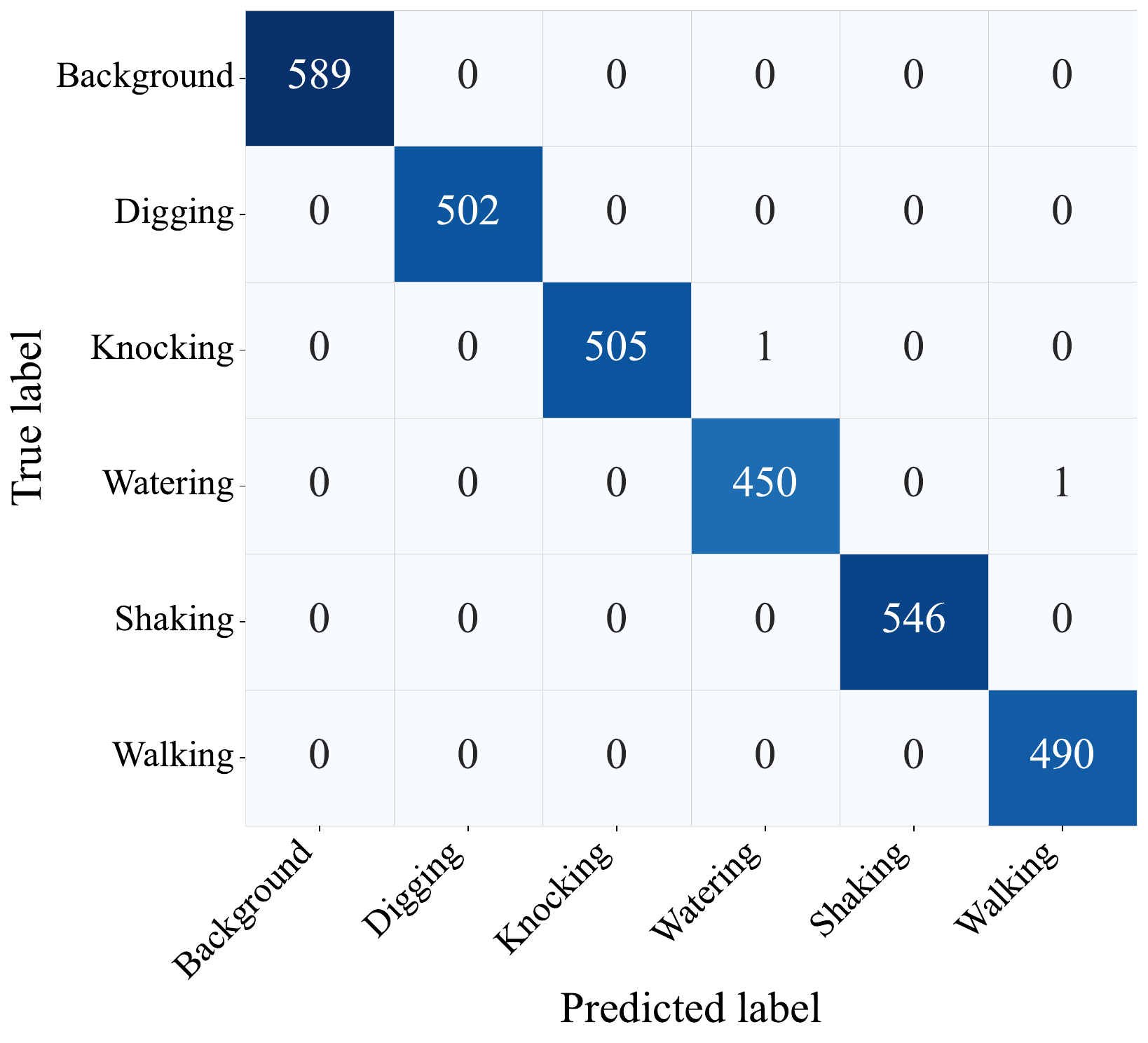} \\
        (a)
    \end{minipage}
    \hfill  
    \begin{minipage}{0.48\columnwidth}
        \centering
        \includegraphics[width=\textwidth]{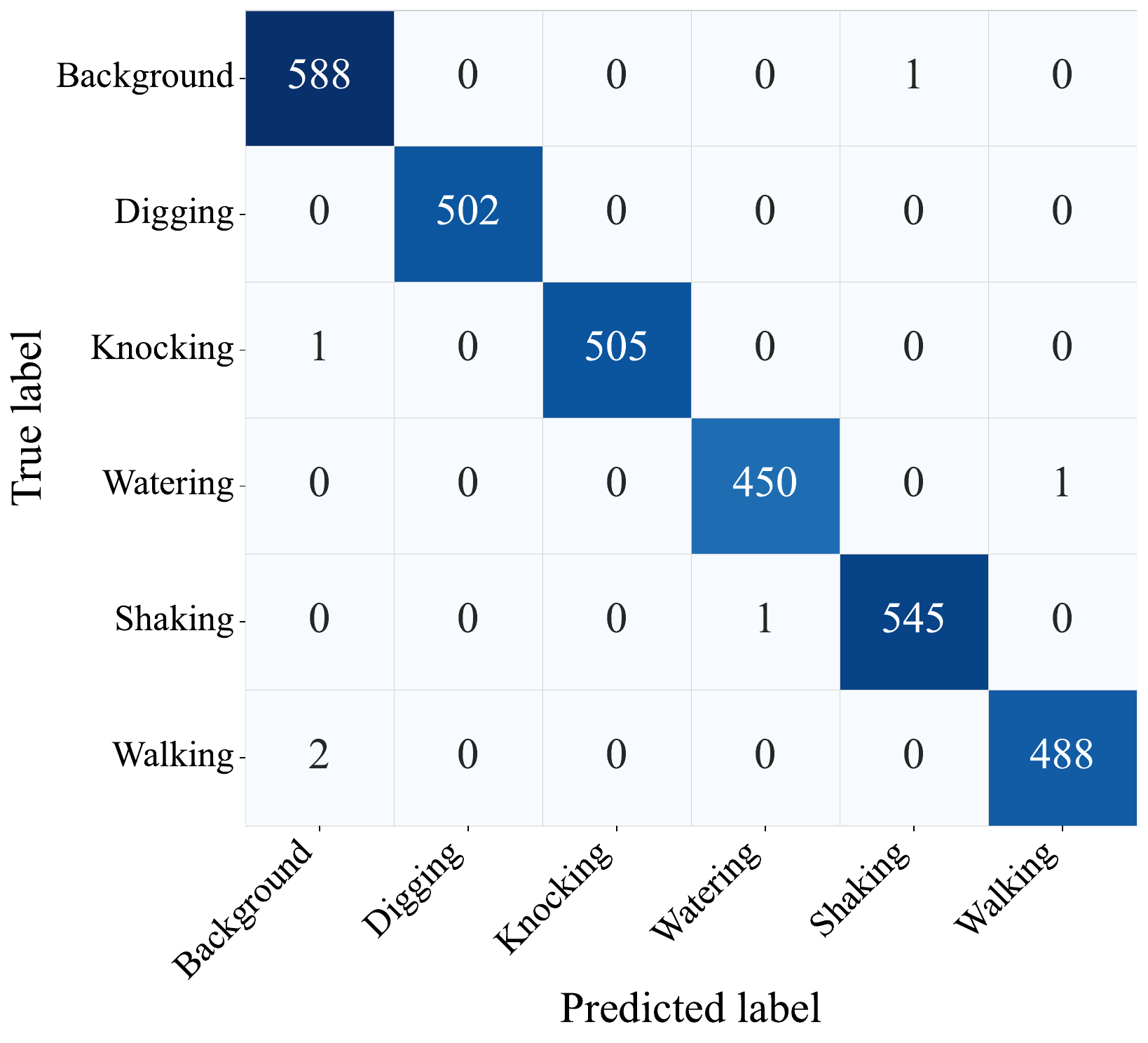} \\
        (b)
    \end{minipage}
    
    \vspace{0.3cm}  
    
    \begin{minipage}{0.48\columnwidth}
        \centering
        \includegraphics[width=\textwidth]{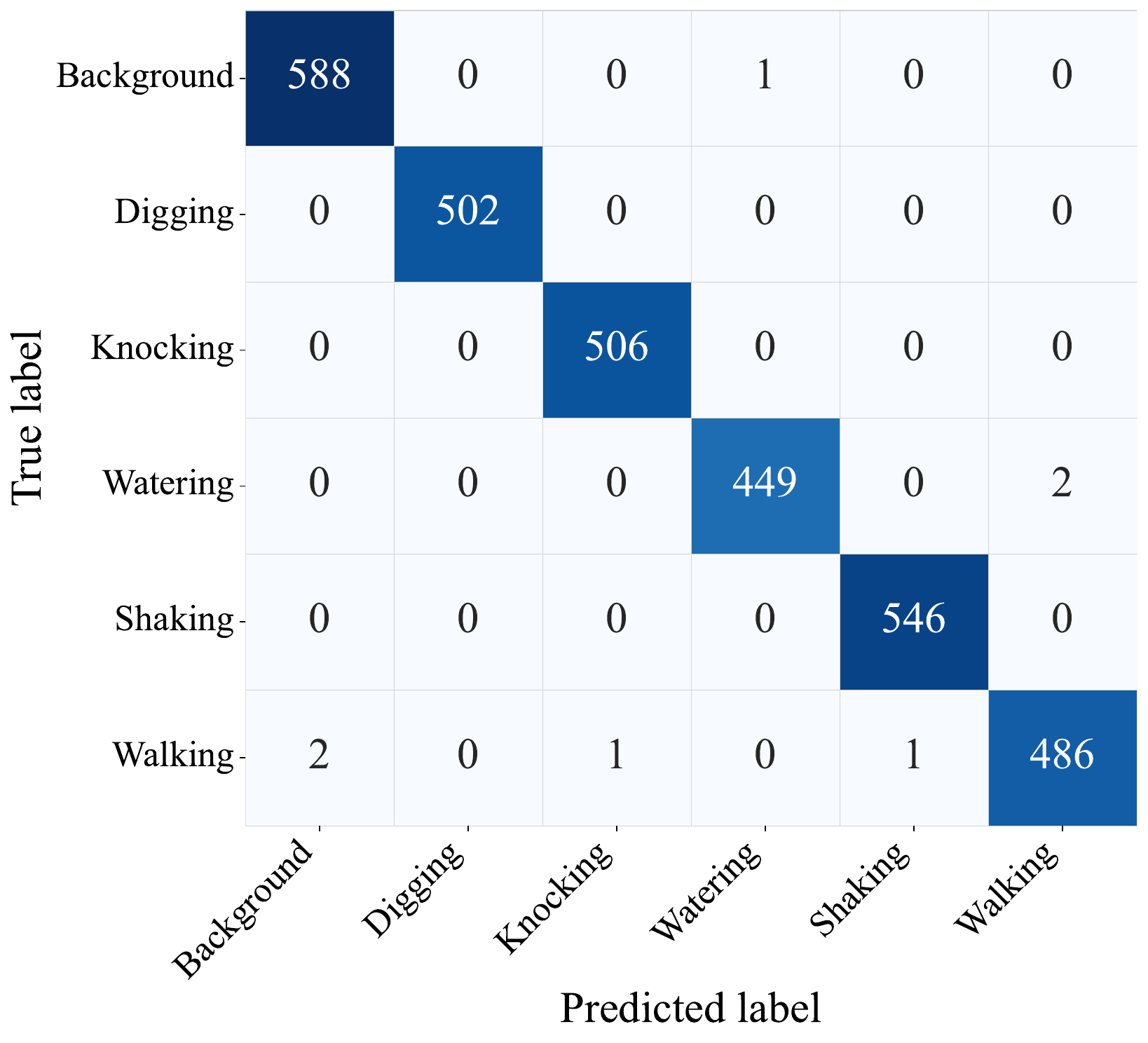} \\
        (c)
    \end{minipage}

    \caption{Confusion matrices for the ablation study. (a) STFT-AECNN (Full Model). (b) Model without SEAM attention. (c) Model without Triplet Loss.}
    \label{fig:ablation_confusion_matrices}
\end{figure}

Fig.~\ref{fig:ablation_confusion_matrices}  visualizes the confusion matrices from the ablation study. The full STFT-AECNN model (Fig.~\ref{fig:ablation_confusion_matrices}(a)), which serves as our baseline, exhibits an almost perfect diagonal with only two misclassifications: one 'knocking' (2) instance confused with 'watering' (3), and one 'watering' (3) with 'walking' (5). This near-perfect performance underscores the model's high discriminative power.

In contrast, both ablated models (Fig.~\ref{fig:ablation_confusion_matrices}(b) and Fig.~\ref{fig:ablation_confusion_matrices}(c)) exhibit a clear degradation in performance, manifested as an increase in scattered off-diagonal errors compared to the full model. This highlights the distinct yet complementary roles of the SEAM attention and Triplet Loss in classifying challenging DAS events. The SEAM module is crucial at the feature extraction stage, enhancing the model's ability to extract discriminative signatures from noisy, multi-channel inputs, especially for spatially sparse events like 'walking'. The Triplet Loss, in turn, operates on these features to build a well-structured embedding space, improving the separation between conceptually distinct but featurally similar categories (e.g., 'watering' vs. 'walking'). Therefore, the increase in misclassifications in both scenarios demonstrates that robust performance requires both mechanisms: one to 'see' the event signal clearly within the noise, and the other to 'understand' and separate its feature representation. The synergy is essential.
    
\subsection{Comparative Analysis of Classification Performance}
To rigorously evaluate the performance of our proposed STFT-AECNN architecture, we conducted a comprehensive comparison against a diverse set of models on the public BJTU $\Phi$-OTDR dataset. For all baseline models, we faithfully reproduced their architectures according to the descriptions in their original papers to establish a fair and reliable benchmark. The rationale for selecting these models is as follows:
Firstly, we benchmark against an SVM \cite{cao2023open} to justify the adoption of a deep learning approach over traditional methods reliant on handcrafted features.
Secondly, we selected a standard 2D-CNN \cite{cao2023open} as a baseline. Since our STFT-AECNN model is an enhanced 2D-CNN architecture using a novel data representation, this comparison allows us to verify the combined performance gains from our two core innovations: the proposed STFT preprocessing and the SEAM attention module.
Finally, we included ST-T \cite{jiang2024st}  and ViT-VSEC \cite{jiang2024high} to benchmark our work against representative Transformer architectures. ST-T is a SpatioTemporal Transformer that partitions multi-location time series data along the time axis, aiming to capture both temporal and spatial features simultaneously. ViT-VSEC, on the other hand, adopts a computer vision approach, treating the entire spatiotemporal data matrix as an image for patch-based processing with a Vision Transformer. With their powerful self-attention mechanisms, Transformers exhibit exceptional capabilities in capturing long-range signal dependencies; however, this capability typically comes with significant computational overhead and model complexity.

By comparing our model against these  different classes of architectures, we aim to demonstrate that our proposed method not only outperforms its foundational architecture but can also rival top-performing Transformer models while offering a superior balance of performance and efficiency. The detailed metrics in Table~\ref{tab:performance_comparison_final} unequivocally support these claims. Firstly, our STFT-AECNN achieves a state-of-the-art peak accuracy of 99.94\%, representing a substantial leap of over ten percentage points against the traditional SVM and over six percentage points against the foundational 2D-CNN. Secondly, and perhaps more importantly, our model surpasses the leading Transformer competitors, ST-T and ViT-VSEC, demonstrating that top-tier accuracy is achievable without their typical computational complexity. Finally, a key differentiator is the model's exceptional consistency; as indicated by F1-scores remaining at or above 99.8\% for all event types, our model provides highly balanced and reliable performance, avoiding the class-specific trade-offs sometimes seen in other architectures.

\begin{figure*}[htbp]
    \centering

    \begin{subfigure}[t]{0.32\textwidth}
        \centering
        \includegraphics[width=\linewidth]{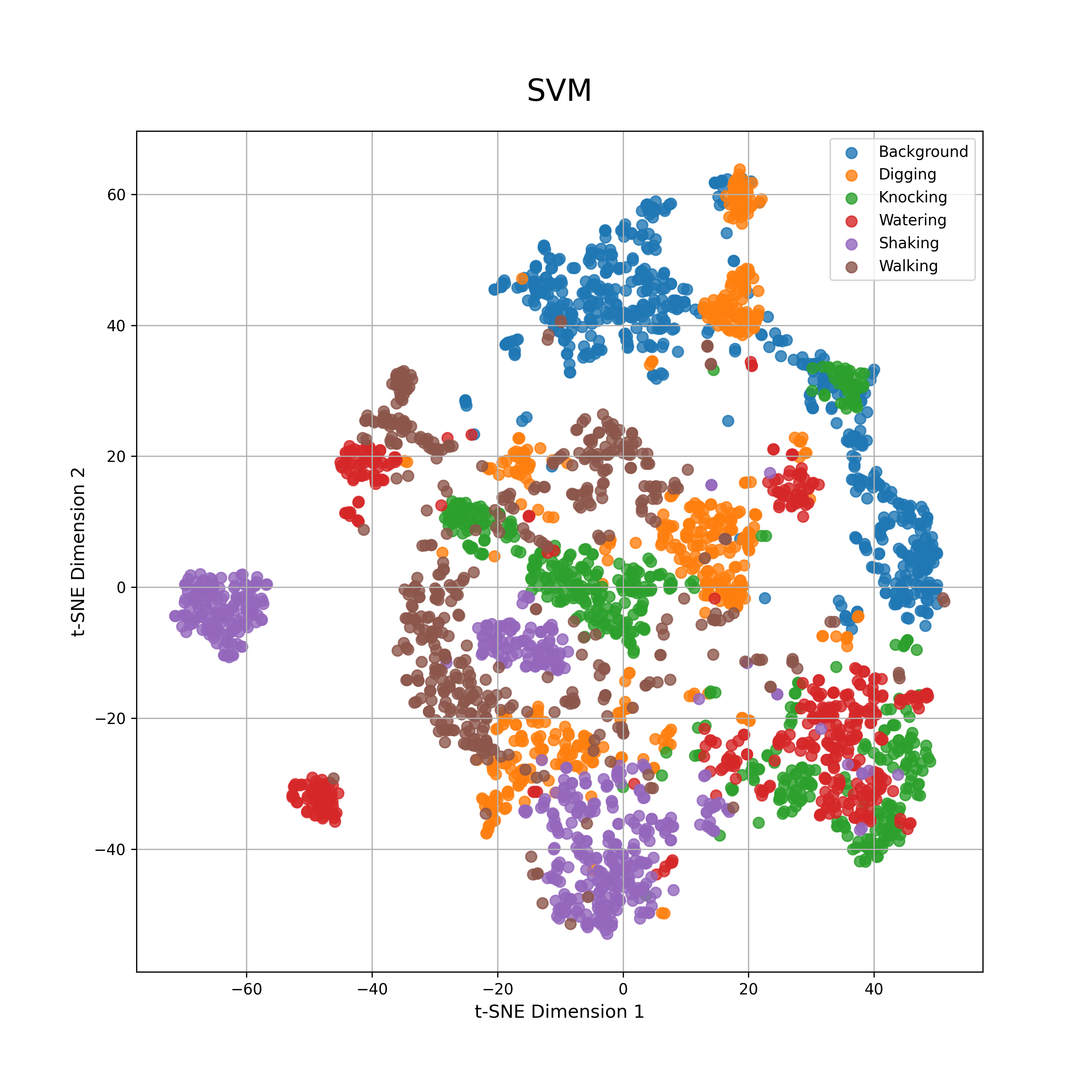}
        \caption{SVM}
        \label{fig:svm_tsne}
    \end{subfigure}
    \hfill 
    \begin{subfigure}[t]{0.32\textwidth}
        \centering
        \includegraphics[width=\linewidth]{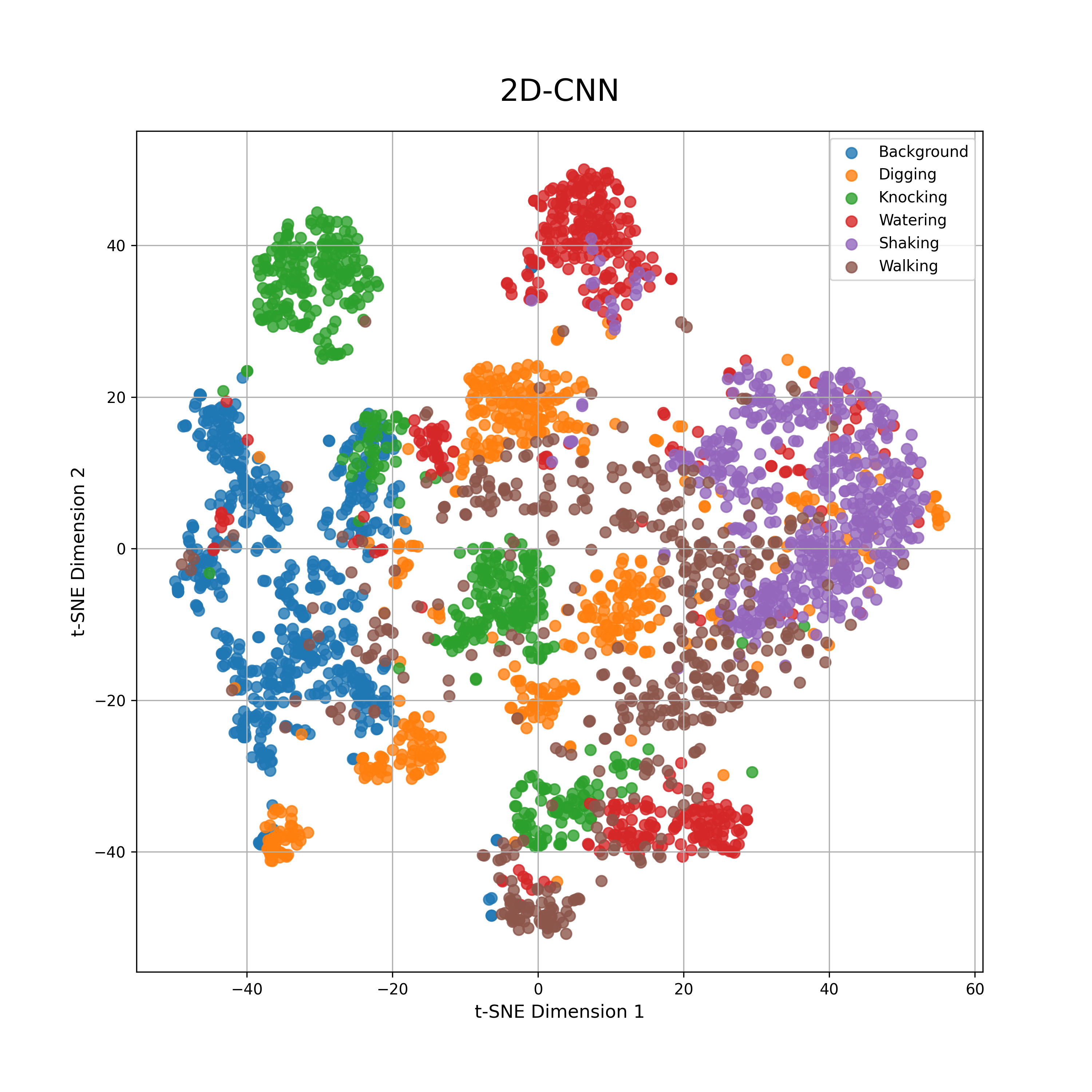}
        \caption{2D-CNN}
        \label{fig:2dcnn_tsne}
    \end{subfigure}
    \hfill 
    \begin{subfigure}[t]{0.32\textwidth}
        \centering
        \includegraphics[width=\linewidth]{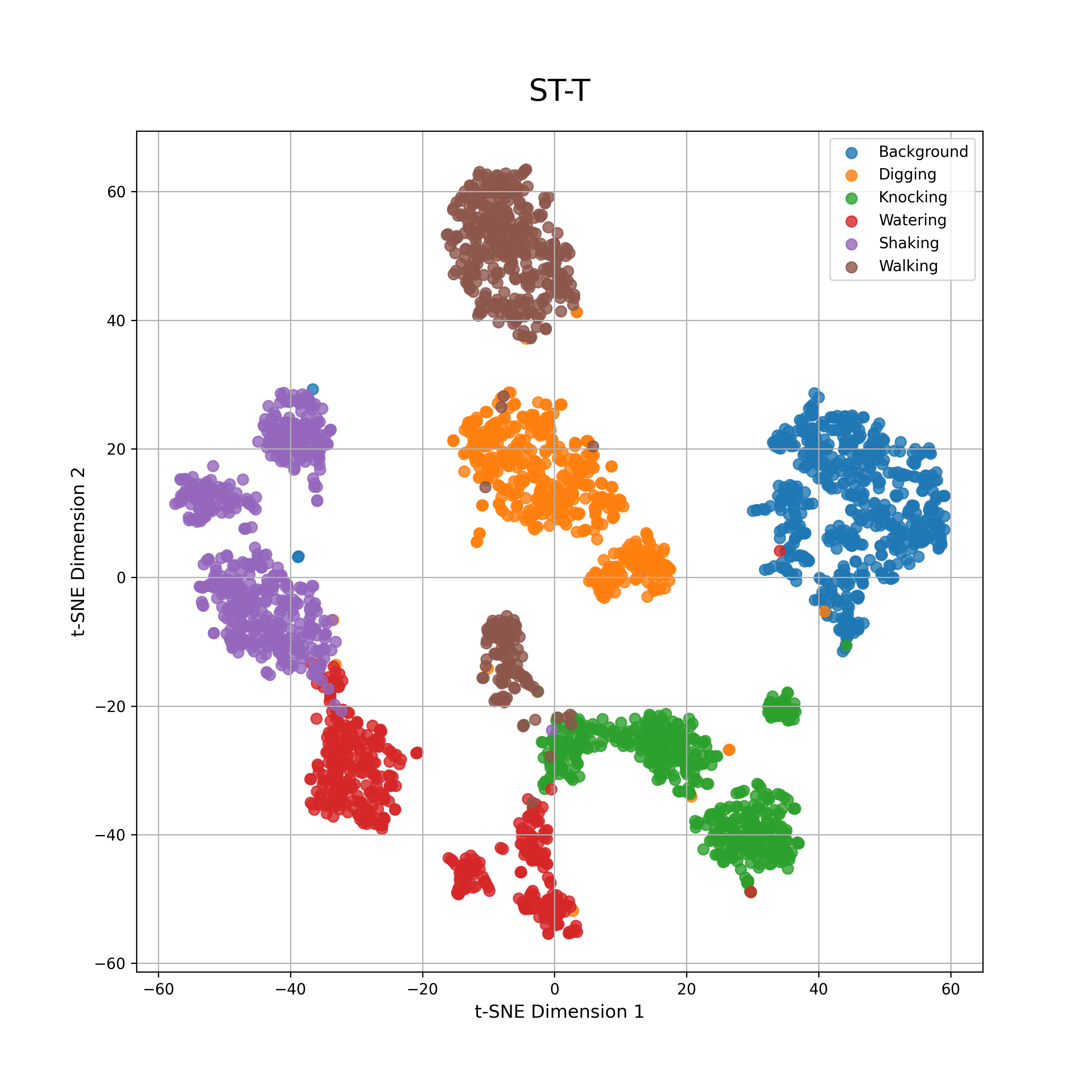}
        \caption{ST-T}
        \label{fig:stt_tsne}
    \end{subfigure}

    \vspace{1em} 

    \begin{minipage}{0.65\textwidth} 
        \centering 
        \begin{subfigure}[t]{0.48\linewidth} 
            \centering
            \includegraphics[width=\linewidth]{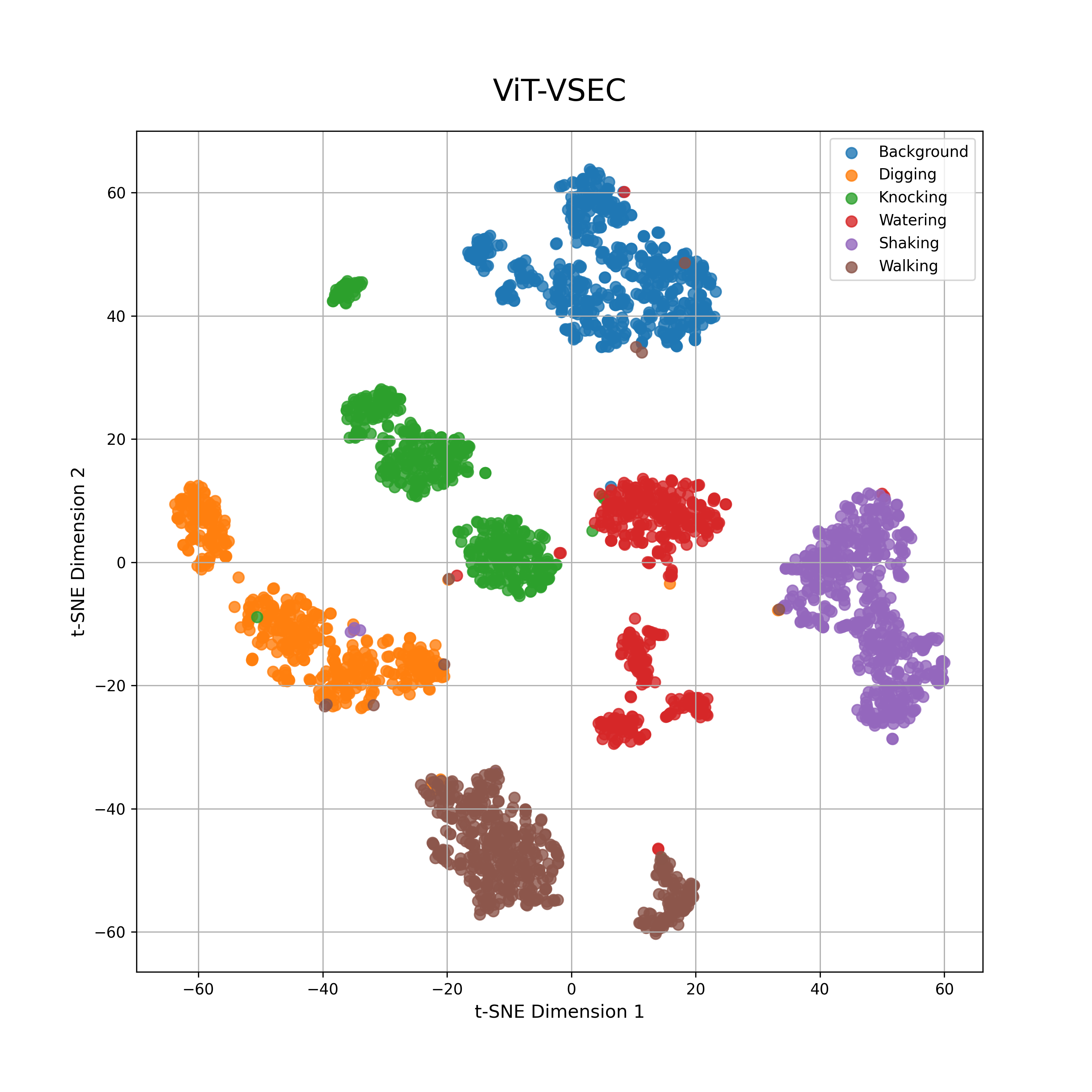}
            \caption{ViT-VSEC}
            \label{fig:vit_vsec_tsne}
        \end{subfigure}
        \hfill 
        \begin{subfigure}[t]{0.48\linewidth}
            \centering
            \includegraphics[width=\linewidth]{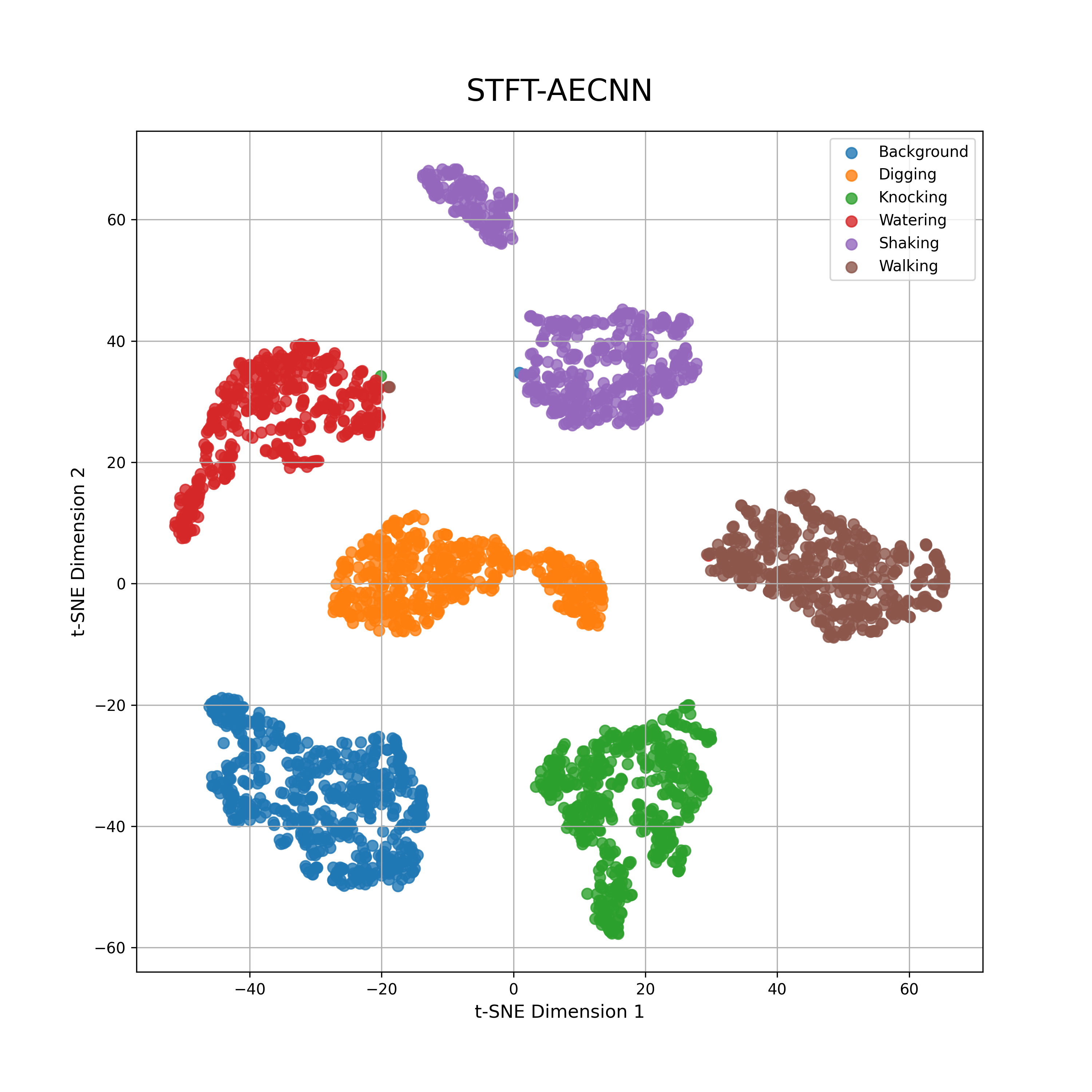}
            \caption{STFT-AECNN (Ours)}
            \label{fig:stft_aecnn_tsne}
        \end{subfigure}
    \end{minipage}

    \caption{Comparison of learned feature spaces, visualizing the superior class separability achieved by STFT-AECNN. Each subplot shows the t-SNE embedding of test set features for a different model: (a) SVM, (b) 2D-CNN, (c) ST-T, (d) ViT-VSEC, and (e) our proposed STFT-AECNN.}
    \label{fig:tsne_visualization}
\end{figure*}

\begin{table}[htbp]
    \centering
    \caption{Per-class Classification Performance Comparison}
    \label{tab:performance_comparison_final}
    \begin{tabular}{llccc}
        \toprule
        \textbf{Model} & \textbf{Event Type} & \textbf{Prec.(\%)} & \textbf{Rec.(\%)} & \textbf{F1(\%)} \\
        \midrule
        \multirow{7}{*}{\shortstack[l]{STFT-AECNN  \\ \textbf{Acc=99.94\%}}} 
        & Background (0) & \textbf{100.0} & \textbf{100.0} & \textbf{100.0} \\
        & Digging (1)    & \textbf{100.0} & \textbf{100.0} & \textbf{100.0} \\
        & Knocking (2)   & \textbf{100.0} & \textbf{99.8}  & \textbf{99.9}  \\
        & Watering (3)   & \textbf{99.8}  & \textbf{99.8}  & \textbf{99.8}  \\
        & Shaking (4)    & \textbf{100.0} & \textbf{100.0} & \textbf{100.0} \\
        & Walking (5)    & \textbf{99.8}  & \textbf{100.0} & \textbf{99.9}  \\
        \cmidrule(lr){2-5}
        & Average & \textbf{99.9} & \textbf{99.9} & \textbf{99.9} \\
        \midrule
        
        \multirow{7}{*}{\shortstack[l]{ViT-VSEC  \\ Acc=98.99\%}} 
        & Background (0) & 99.5  & 99.8  & 99.7  \\
        & Digging (1)    & 97.8  & 99.0  & 98.4  \\
        & Knocking (2)   & 99.4  & 99.0  & 99.2  \\
        & Watering (3)   & 98.9  & 98.9  & 98.9  \\
        & Shaking (4)    & 99.3  & 99.3  & 99.3  \\
        & Walking (5)    & 99.0  & 97.8  & 98.4  \\
        \cmidrule(lr){2-5}
        & Average & 99.0 & 99.0 & 99.0 \\
        \midrule
        
        \multirow{7}{*}{\shortstack[l]{ST-T  \\ Acc=97.99\%}}
        & Background (0) & 99.8  & 98.1  & 99.0  \\
        & Digging (1)    & 97.4  & 98.2  & 97.8  \\
        & Knocking (2)   & 97.3  & \textbf{99.8}  & 98.5  \\
        & Watering (3)   & 97.7  & 95.3  & 96.5  \\
        & Shaking (4)    & 96.6  & 99.3  & 97.9  \\
        & Walking (5)    & 99.0  & 96.7  & 97.8  \\
        \cmidrule(lr){2-5}
        & Average & 98.0 & 97.9 & 97.9 \\
        \midrule
        
        \multirow{7}{*}{\shortstack[l]{2D-CNN  \\ Acc=93.74\%}} 
        & Background (0) & 96.9  & 96.4  & 96.7  \\
        & Digging (1)    & 94.2  & 94.6  & 94.4  \\
        & Knocking (2)   & 93.1  & 96.4  & 94.8  \\
        & Watering (3)   & 91.6  & 91.4  & 91.5  \\
        & Shaking (4)    & 92.8  & 96.9  & 94.8  \\
        & Walking (5)    & 93.1  & 85.5  & 89.1  \\
        \cmidrule(lr){2-5}
        & Average & 93.6 & 93.5 & 93.6 \\
        \midrule
        
        \multirow{7}{*}{\shortstack[l]{SVM \\ Acc=89.88\%}} 
        & Background (0) & 89.1  & 98.1  & 93.4  \\
        & Digging (1)    & 85.9  & 88.6  & 87.3  \\
        & Knocking (2)   & 91.9  & 87.4  & 89.6  \\
        & Watering (3)   & 92.8  & 85.1  & 88.8  \\
        & Shaking (4)    & 91.0  & 96.5  & 93.7  \\
        & Walking (5)    & 89.4  & 80.8  & 84.9  \\
        \cmidrule(lr){2-5}
        & Average & 90.0 & 89.4 & 89.6 \\
        \bottomrule
    \end{tabular}
\end{table}
This superior performance, as detailed in Table~\ref{tab:performance_comparison_final}, is fundamentally rooted in the quality of the feature representations learned by the model. To intuitively verify this, we visualized the high-dimensional features extracted from the test set using t-SNE, as presented in Fig.~\ref{fig:tsne_visualization}.
This visualization clearly illustrates the hierarchy of discriminative power among the models. The plot for our STFT-AECNN model exhibits an exceptionally well-organized feature space, where each event category forms a dense, compact, and distinctly separated cluster. In stark contrast, the feature spaces of the SVM and 2D-CNN show significant class overlap and confusion, visually corroborating their lower performance. While the Transformer-based models, ST-T and ViT-VSEC, produce much cleaner clusters than the SVM and 2D-CNN, their intra-class compactness and inter-class separability are still noticeably inferior to those of our model.
Therefore, this visual analysis provides compelling evidence that by synergizing STFT preprocessing with an attention-enhanced CNN, our method learns a superior feature representation, which is the fundamental reason for its state-of-the-art accuracy and robustness.

	\subsection{Analysis of Efficiency and Practical Viability}
\begin{figure}[H]
\centering
\includegraphics[width=\columnwidth]{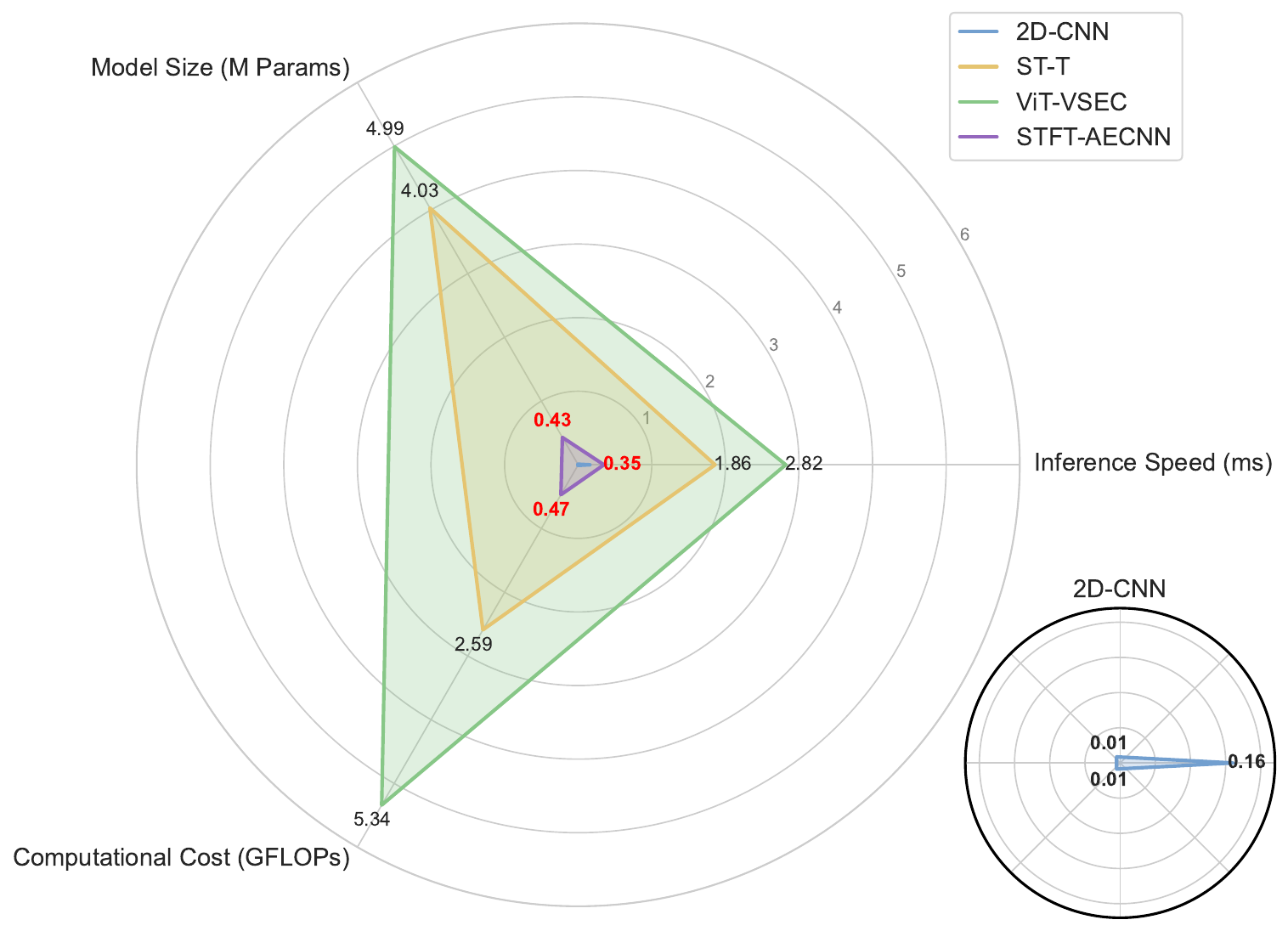}
\caption{Radar chart comparing model efficiency across key metrics: Model Size (M params), Inference Speed (ms), and Computational Cost (GFLOPs). The 2D-CNN model is plotted on a separate, magnified scale in the bottom right for clarity.}
\label{fig:model_performance_comparison_v3}
\end{figure}

Beyond accuracy, this section evaluates model efficiency and practical viability through a series of key metrics. The evaluation centers on three critical metrics for edge deployment: inference speed, which dictates real-time responsiveness; computational cost, which reflects the required processing power; and model size, which impacts memory and storage constraints. With these criteria established, we first narrow the scope of our efficiency comparison by addressing the baselines. The SVM is excluded for two reasons: its fundamentally different evaluation methodology makes direct efficiency comparisons (e.g., in FLOPs or training time per epoch) infeasible, and its low accuracy is insufficient for the high-reliability requirements of edge-based $\Phi$-OTDR  systems. Similarly, while the baseline 2D-CNN demonstrates highly competitive efficiency metrics (as shown in Fig.~\ref{fig:model_performance_comparison_v3}), it is also impractical for deployment due to its lower accuracy. Therefore, our subsequent analysis focuses on the crucial efficiency trade-offs among the high-accuracy models: our proposed STFT-AECNN and the Transformer-based competitors.

This detailed analysis, illustrated in the comprehensive efficiency comparison in Fig.~\ref{fig:model_performance_comparison_v3}, reveals that our STFT-AECNN establishes a clear superiority in practical deployment. This advantage is not marginal; it represents a fundamental leap in operational feasibility for edge-based DAS.
Specifically, its inference time of just 0.35 ms is over 5 times faster than ST-T (1.86 ms) and over 8 times faster than ViT-VSEC (2.82 ms). Such a significant speed advantage makes our model uniquely suited for time-sensitive IoT applications, directly addressing the high-latency bottleneck that renders many powerful Transformer-based architectures impractical at the edge.
Furthermore, our model’s resource footprint is a fraction of the Transformers’. It requires less than 20\% of the computational cost (FLOPs) of ST-T and 10\% of ViT-VSEC. Similarly, its model size is exceptionally compact at only 0.43M parameters, approximately 10\% of either Transformer model. This dual efficiency in computation and memory makes STFT-AECNN ideal for deployment on resource-constrained edge devices, overcoming the substantial hardware barriers presented by larger, more demanding architectures. 

The preceding analysis establishes the deployment-time superiority of our model in terms of inference speed and resource consumption. However, the practical viability of a system, particularly in the IoT context, extends beyond inference performance alone. It also encompasses the model's entire lifecycle, including initial development, iterative tuning, and long-term maintenance. In this lifecycle, high training efficiency is paramount. It dramatically shortens development cycles, enables rapid experimentation with different hyperparameters, and reduces the overall cost and time required to deploy models across diverse edge hardware. Thus, an examination of training efficiency is equally important for assessing a model's practical utility.
 
\begin{figure}[H]
    \centering
    \includegraphics[width=\columnwidth]{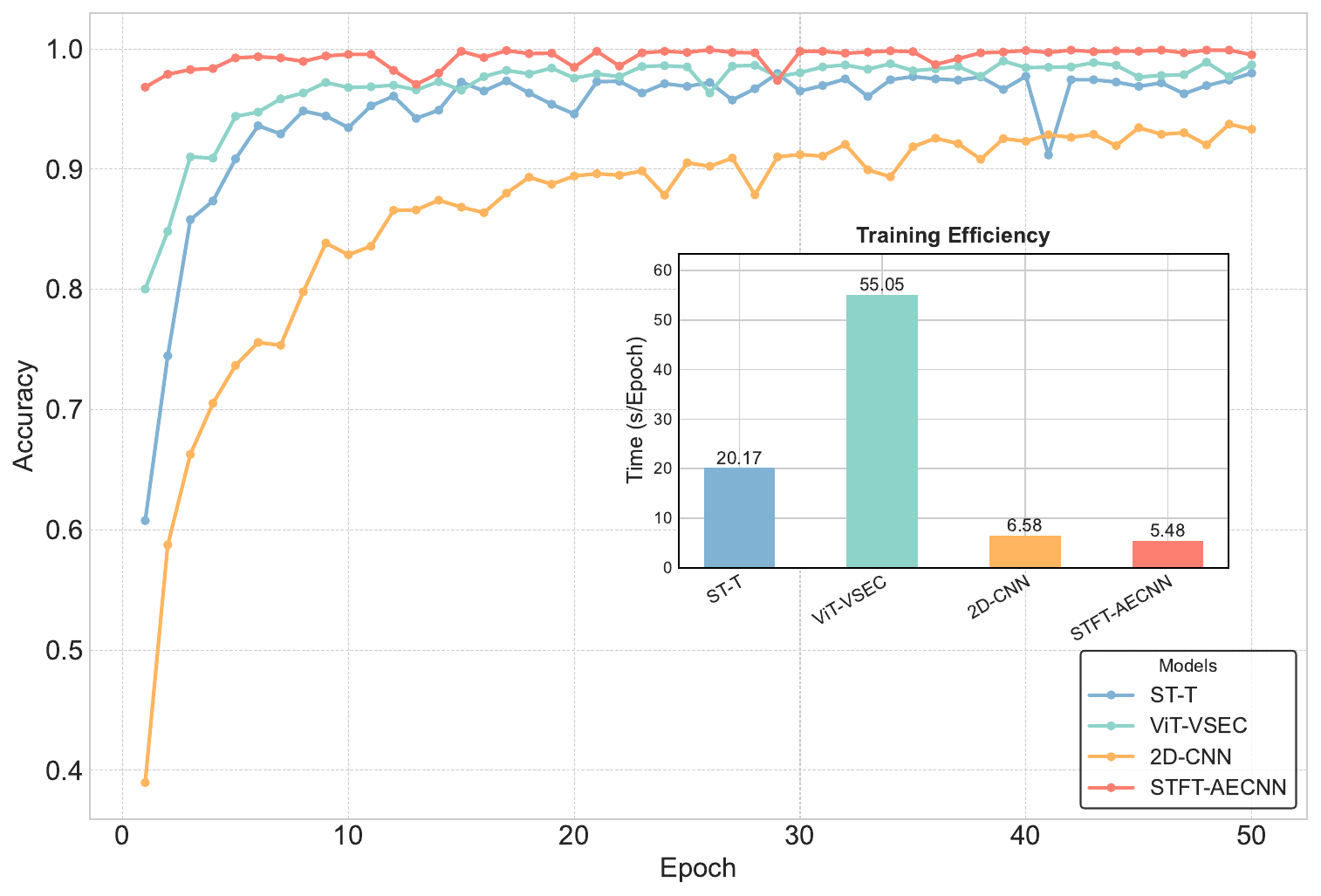}
    \caption{Convergence dynamics and performance comparison of STFT-AECNN and baseline models. The main plot tracks the test accuracy over epochs, while the inset shows the training time per epoch.}
    \label{fig:training_curves}
\end{figure}

This per-epoch speed advantage is further amplified by our model’s superior learning dynamics, as revealed by the convergence curves in Fig.~\ref{fig:training_curves}. Our model not only processes each epoch faster, but it also extracts more value from each one. This is immediately evident as its performance curve remains almost consistently superior to all baselines, including the high-performing Transformer models. Furthermore, it demonstrates significantly more efficient convergence, rapidly ascending to over 95\% accuracy within the first few epochs and stabilizing at a superior performance level above 99\% with minimal volatility. In contrast, the Transformer models require more epochs to approach their peak accuracy and exhibit greater performance fluctuations, while the 2D-CNN lags substantially.
Moreover, in terms of long-term operational efficiency, training speed emerges as a decisive factor. As demonstrated by the inset bar chart in Fig.~\ref{fig:training_curves}, our model’s training is remarkably swift, even surpassing the baseline 2D-CNN. Its time per epoch is merely 10\% of that required by ViT-VSEC and only 27\% of that of ST-T. The primary advantage of the model's rapid training lies in its ability to dramatically shorten the development cycle, a critical factor when tuning models for diverse and resource-constrained edge hardware. More importantly, this efficiency reduces the operational overhead of model maintenance, enabling agile deployments with rapid updates and minimal downtime. This dual advantage—faster processing per epoch and more effective learning within each epoch—translates to a dramatically more efficient overall training process, which is a critical asset for practical deployment and iterative development at the edge.

\section{CONCLUSIONS}\label{conclusion}
In this paper, we proposed a lightweight network , STFT-AECNN, to address the critical challenge of achieving both state-of-the-art accuracy and high computational efficiency for $\Phi$-OTDR event classification in resource-constrained IoT environments. Specifically, we introduced a STFT-based data representation that transforms raw multi-channel signals into stacked spectrograms, preserving the signal's inherent spatiotemporal structure while enabling efficient 2D-CNN processing. Furthermore, our design incorporates a custom SEAM attention module and a joint loss function to enhance discriminative feature learning, enabling the model to adaptively focus on sparse event signatures. Extensive experiments demonstrated that our approach not only surpasses foundational architectures but also rivals more complex, state-of-the-art methods in accuracy, all while maintaining a minimal computational footprint. These results underscore the potential of STFT-AECNN as a practical and scalable solution, paving the way for real-time intelligence at the edge across IoT sensing systems.



\end{document}